\documentstyle[preprint,aps]{revtex}
\begin{document}
\draft
\preprint{titcmt-95-10, adap-org/9504004}
\title{Numerical investigation of
surface level instability due to tube in vibrating bed of powder}
\author{Yasushi Maeno\footnote{All correspondence should be sent to
Y-h. Taguchi, at the same address as above.
Electronic Address: ytaguchi@cc.titech.ac.jp}}
\address{c/o Dr. Y-h. Taguchi, Department of Physics, Tokyo Institute of
Technology\\
Oh-okayama, Meguro-ku, Tokyo 152, Japan}
\date{\today}
\maketitle
\begin{abstract}
Surface level instability when tube is injected into vibrating bed of powder,
which was originally found in experiments,
is investigated numerically.
We find that thicker (thiner) tube makes surface level inside tube
higher (lower) than surface level outside tube.
With fixed acceleration amplitude of vibration,
surface level inside tube becomes higher as
amplitude of vibration increases, which
can be explained by
considering the dependence upon strength of convective flow.
\end{abstract}

\section{Introduction}

The dynamics of granular material attracted many attentions of
physicists\cite{review}.
Among them, vibrating bed of powder was
studied by Faraday half and a century ago\cite{Faraday},
and many papers were written about its behavior.
This is because vibrating bed can exhibit
many interesting phenomena like
convection\cite{conv}, surface heaping\cite{heap},
surface fluidization\cite{surface},
size segregation\cite{seg}, and turbulence\cite{turb}.
Other than those phenomena,
surface level instability due to injection of tube into
vibrating bed of powder was observed in experiments\cite{akiyama}.
The main purpose of this paper is
to reproduce this phenomenon numerically.

The organization of this paper is as follows.
In Sec. \ref{sec:exp}, we briefly
summarized experimental findings by Akiyama and Shimomura.
Numerical modeling used in this paper is explained in Sec. \ref{sec:num}
and results will be presented in Sec. \ref{sec:res}.
Summary and discussion can be found in Sec. \ref{sec:sum}.

\section{Experiments}
\label{sec:exp}

Vibrating bed of powder is a vessel
filled with granular matter, typically mono-disperse glass beads,
and vessel is shaken vertically as strong as gravity acceleration.
When acceleration amplitude of vibration exceeds
critical value, which is usually a little bit larger than
gravity acceleration, the bed exhibits several instabilities,
e.g., surface heaping\cite{heap}, surface fluidization\cite{surface},
and convection\cite{conv}.
Further increase of acceleration amplitude results in disappearance of
heap, and surface starts to fluctuate violently.
In order to measure shear friction in this vibrating bed,
Akiyama and Shimomura\cite{akiyama} injected
thick tube into vibrating bed and
found vary surprising effect.
With tube fixed in space, i.e.,
tube does not vibrate,
the surface level inside tube differs from
surface level outside tube (Fig. \ref{fig:schem}),
even if  heaping is not observed without tube.
Thus, this instability may be different from
heaping instability observed in vibrating bed.
The surface level difference depends upon
several physical parameters as
diameter of tube, acceleration amplitude,
and particle diameter.
It was rather difficult to understand these dependences
because experiment of powder can be affected
by many other fine differences of conditions,
e.g., moisture, temperature, and so on.
Thus, numerical investigation is much
suitable to understand these dependences in detail.

\section{Numerical model}
\label{sec:num}

Although there are many numerical schemes for
investigating dynamics of powder,
we employ here distinct element method (DEM)\cite{cundal}
to reproduce Akiyama's experiments.
In DEM, granular particle is models as
visco-elastic particle,
whose interaction is limited within short range.
Particularly we employ here
non-spherical models introduced by P\"oschel and Buchholtz\cite{PRL.Poeschel},
which is known to reproduce static friction effect better than
conventional models.
In their non-spherical model,
each granular particle is modeled as
a
set of five sub-particles (See Fig. \ref{fig:non-spherical}).
Although tangential force is ignored,
static friction can be considered as interaction among
surrounding four small sub-particles.
Since relative position between center sub-particle and
surrounding small sub-particles are considered,
rotation of non-spherical particle can be considered effectively.
Each sub-particle
obeys the following equation
\def\r{{\vec r}}
\def\F{{\vec F}}
\begin{equation}
m_i \frac{d^2 \r_i}{dt^2} = \sum_j \F^{ij}_{inner} +
\sum_j\F^{ij}_{outer} \times \Theta ( \ell'_{ij}-\mid \r_j-\r_i\mid)
\end{equation}
where $F^{ij}_{inner}$ and $F^{ij}_{outer}$ are
interaction within a non-spherical particle and
interaction between non-spherical particles respectively.
Here subscript $i$ is $c$ or $r$ depending upon
whether $i$ sub-particle is center (c) sub-particle or
surrounding (r) sub-particle.
$m_i$ is mass of $i$ sub-particle and $\r_i$ is position vector of
$i$ sub-particle.
$\Theta$ is step function, and
$\ell'_{ij}$ is distance between sub-particles which belong to
different non-spherical particles,
\begin{equation}
\l'_{ij}= \left \{
\begin{array}{lc}
2R_r & (i,j=r,r)\\
2R_c & (i,j=c,c)\\
R_c+R_r & (i,j=c,r)\\
\end{array}
\right.
\end{equation}
where $R_i$ is radius of $i$ sub-particle.
$\F^{ij}_{inner}$ is defined as,
\def\n{{\vec n}}
\begin{equation}
\F^{ij}_{inner} = \left [ -k
(\mid \r_i-\r-j\mid-\ell_{ij})
- - -\gamma \left( \frac{d\r_j}{dt}-\frac{d\r_i}{dt} \right )
\cdot \n_{ij} \right ]\n_{ij}
\end{equation}
with $\n_{ij} \equiv (\r_j-\r_i)/\mid \r_j-\r_i\mid$.
$k$ and $\gamma$ are elastic and viscosity constant respectively,
which are related to coefficient of restitution,
$e = \exp \left( - \frac{ \pi \gamma}
{2 \sqrt{k-(\gamma/2)^2}} \right )$.
$\ell_{ij}$ is distance between sub-particles among a
non-spherical particle,
\begin{equation}
\ell_{ij} = \left \{
\begin{array}{lc}
R_c + R_r & (i,j =c,r) \\
\sqrt{2} (R_c + R_r) & (i,j =r,r) \\
\end{array}
\right.
\end{equation}
This means that five sub-particles have interaction only when
their relative positions deviate from that shown in
Fig.\ref{fig:non-spherical}.
$\F_{outer}^{ij}$ is defined as
\begin{equation}
\F_{outer}^{ij} = \left [
- - -k ( \mid \r_i -\r_j \mid -\ell'_{ij} )
- - -\gamma \left ( \frac{d \r_j}{dt} -\frac{d\r_i}{dt}
\right ) \cdot \n_{ij} \right ] \n_{ij}
\end{equation}

In order to simulate vibrating bed of powder,
we have to introduce vessel composed of sub-particles.
Interaction between sub-particles which construct
non-spherical particles and sub-particles which construct
vessels employs the same functional form as $\F^{ij}_{outer}$,
but $\ell'_{ij}$ is replaced with
\begin{equation}
\l'_{ij}= \left \{
\begin{array}{lc}
R_c+R_v & (i,j=c,v)\\
R_r+R_v & (i,j=r,v)\\
\end{array}
\right.
\end{equation}
Motion of sub-particles which construct vessel
is not affected by the interaction with
sub-particles which  constructs non-spherical particle,
but follows given motion of vessel itself,
e.g., vibration or static state.
In addition to this, 'solid plate' is added to the bottom
to prevent particles from falling out of the vessel through bottom.
This plate causes vertical force applied to $i$th sub-particle,
\begin{equation}
F^i_{bottom}=k'[z_{bottom}-z_i],
\end{equation}
where $z_{bottom}$ and $z_i$ are vertical components of
bottom and $i$th sub-particle, respectively.

\section{The results}
\label{sec:res}

In the simulations, we employ following
parameters (See Fig.\ref{fig:schem.num}).
Radius of sub-particles are $R_c=3.0, R_r=0.5$, and
$R_v=1.5$. We ignore distribution of radius and
use identical non-spherical particles.
Number of non-spherical particles are 200,
this means, number of sub-particles is 1000.
Interaction parameters are taken as
$k/m_i=600.0$ and $\gamma/m_i=2.0$ independent of
the kind of sub-particles.
Thus coefficient of restitution $e=0.88$.
$k'$ is take to be $2k$.
For size of vessel, its diameter $D_v=90$ and its height
$H_v=135$.
Tube  whose height $H_c=10.5$ is separated from bottom of vessel
by $H_s=45$.
Gravity acceleration is taken to be 9.8.

Control parameters of simulation are
the acceleration amplitude of vibration $\Gamma$,
angular frequency of vibration $\omega$, and
diameter of tube, $D_c$.
Here $\Gamma$ is related with $\omega$ as
$\Gamma= a\omega^2/g$ where $a$ is amplitude of vibration.
The values of these parameters used in simulations are,
$\Gamma=5.0, 6.0,$ and $7.0$,
$\omega=3.0, 4.0,$ and $5.0$,
$D_c=30,24$, and $18$.
Total length of simulation is over $40$ periods.

Figure \ref{fig:snapshot} shows typical snapshot of simulation.
In order to measure surface level differences, we define height of
surface as follows.
For example,  surface height inside tube $z_i$ is taken such that
number of non-spherical particles above $z_i$ is
half as much non-spherical particles as aligned along section of tube.
(For detail, see Fig. \ref{fig:height}.)
The height $z_o$ outside tube is defined in similar way.
Then height difference between inside and outside tube
$\Delta z = z_i-z_o$.
In addition to this,
we measure flow of non-spherical particles within vessel
following ref. \cite{conv}, i.e.,
by dividing vessel into cells as large as non-spherical
particle, and flow is defined as number of particle transport
between cells per period of vibration.

Figure \ref{fig:del z} shows the dependence of $\Delta z$ upon
several parameters, and
Figure \ref{fig:flow} shows flow patterns and strength for
several parameter values.

\section{Summary and discussion}
\label{sec:sum}

\begin{figure}
\caption{Schematics of conclusion}
\label{fig:summary}
\end{figure}
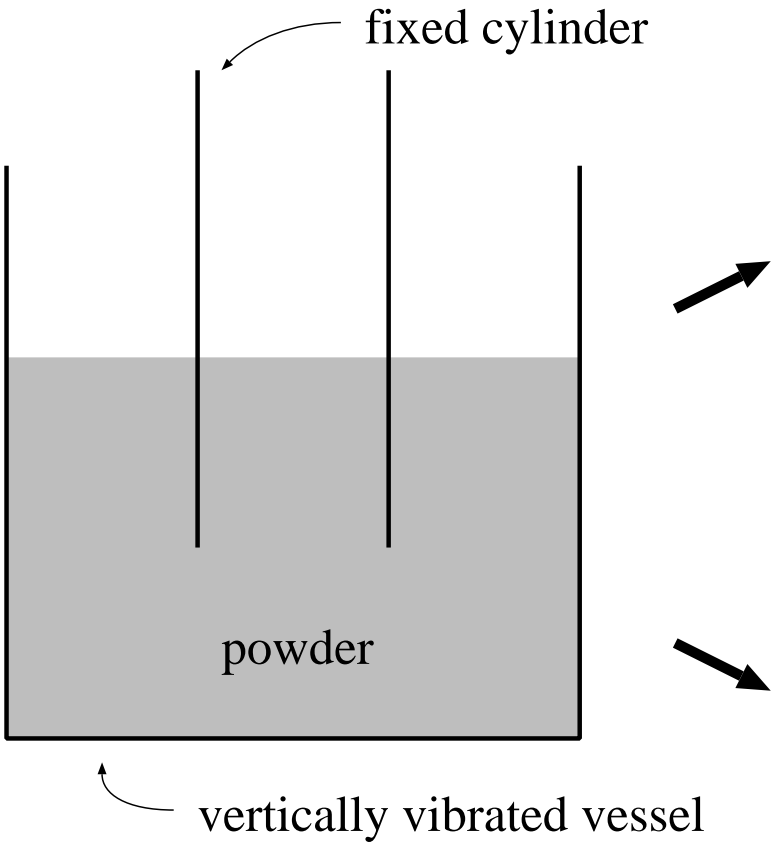
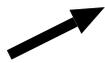
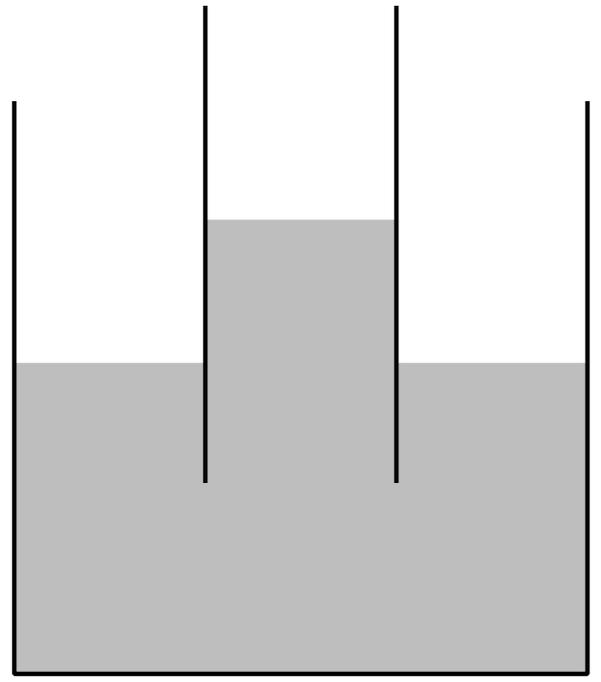
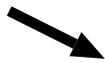
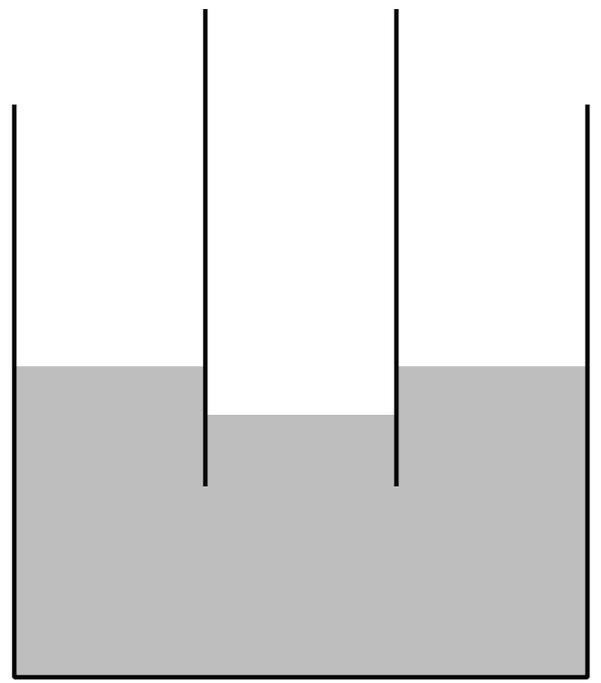
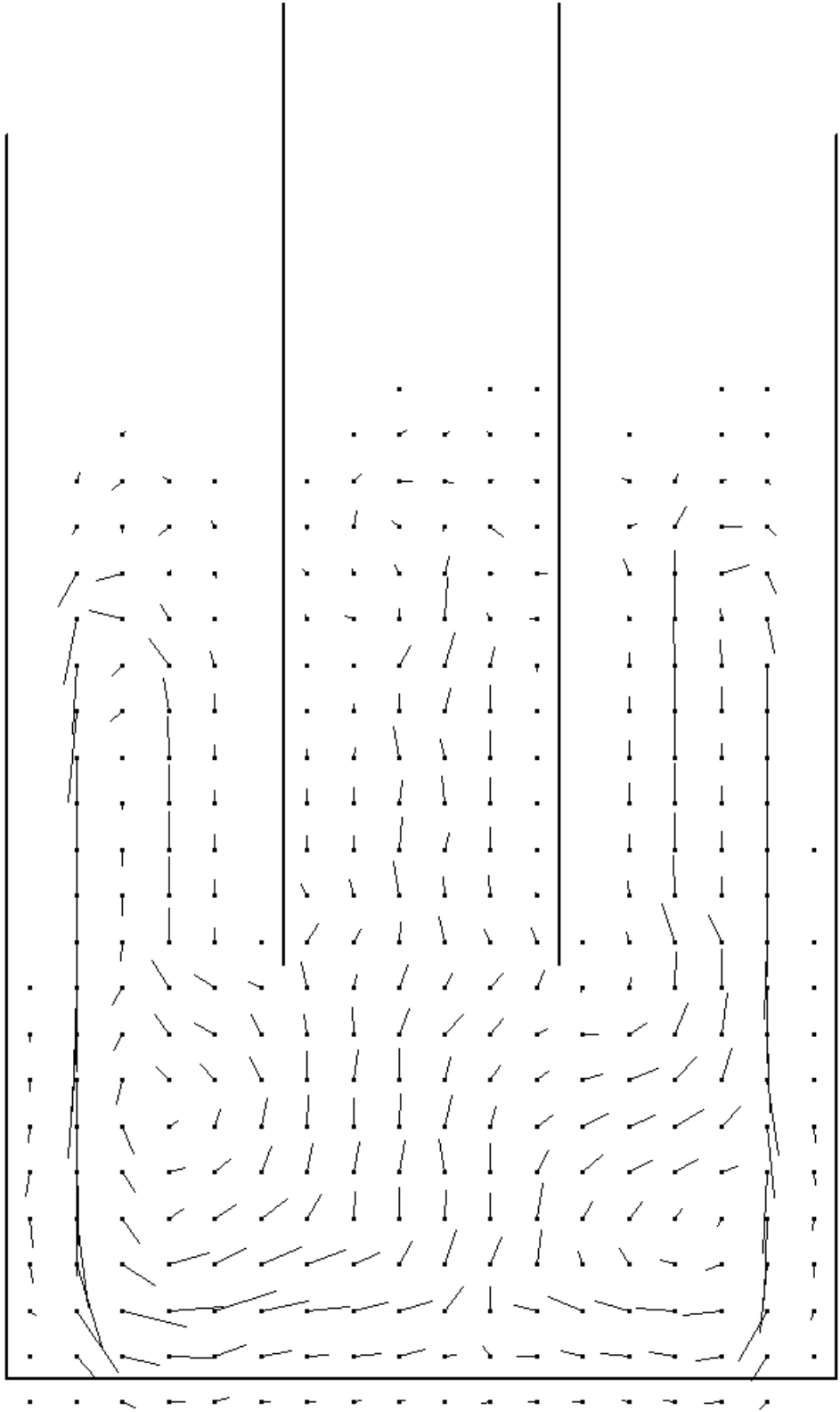
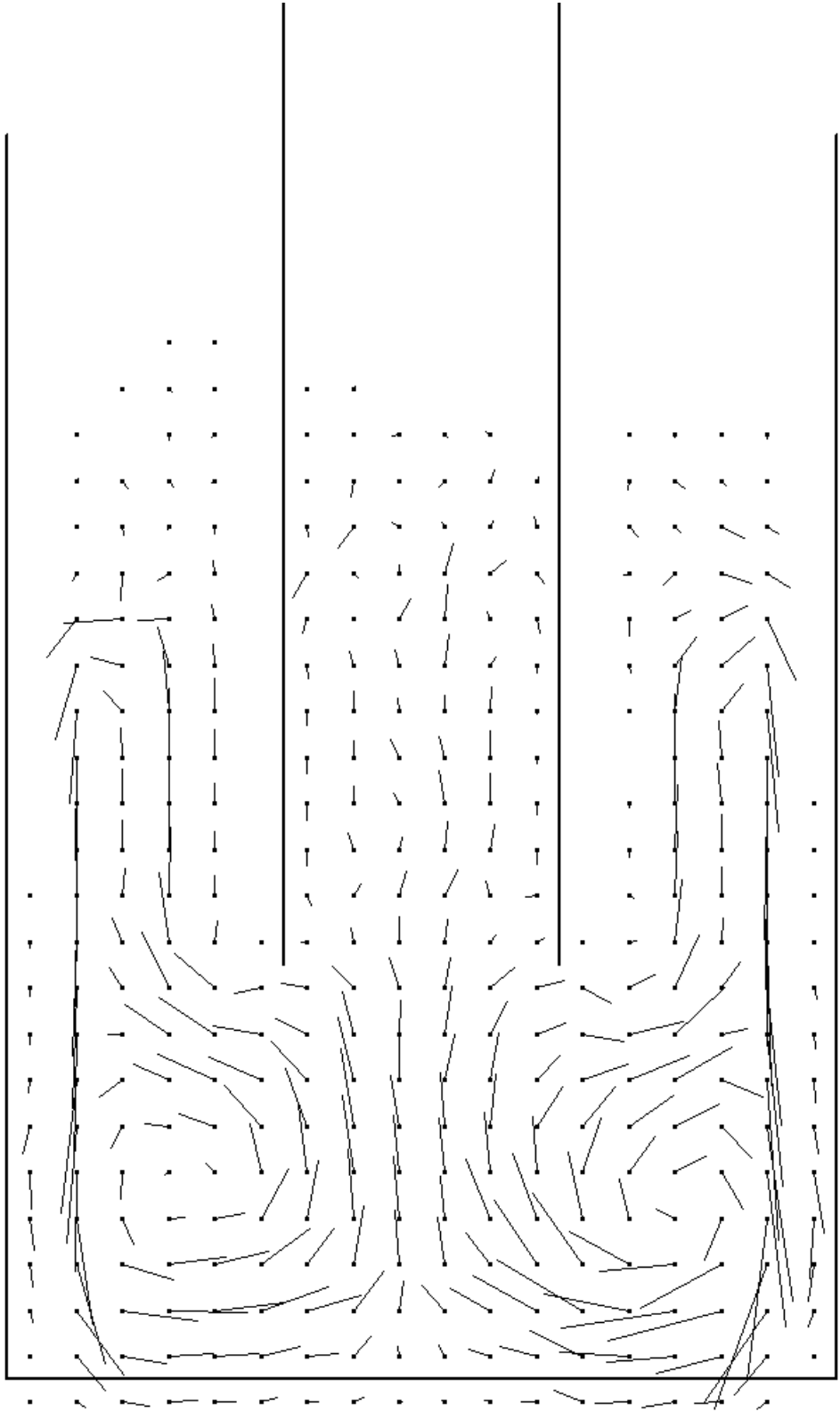
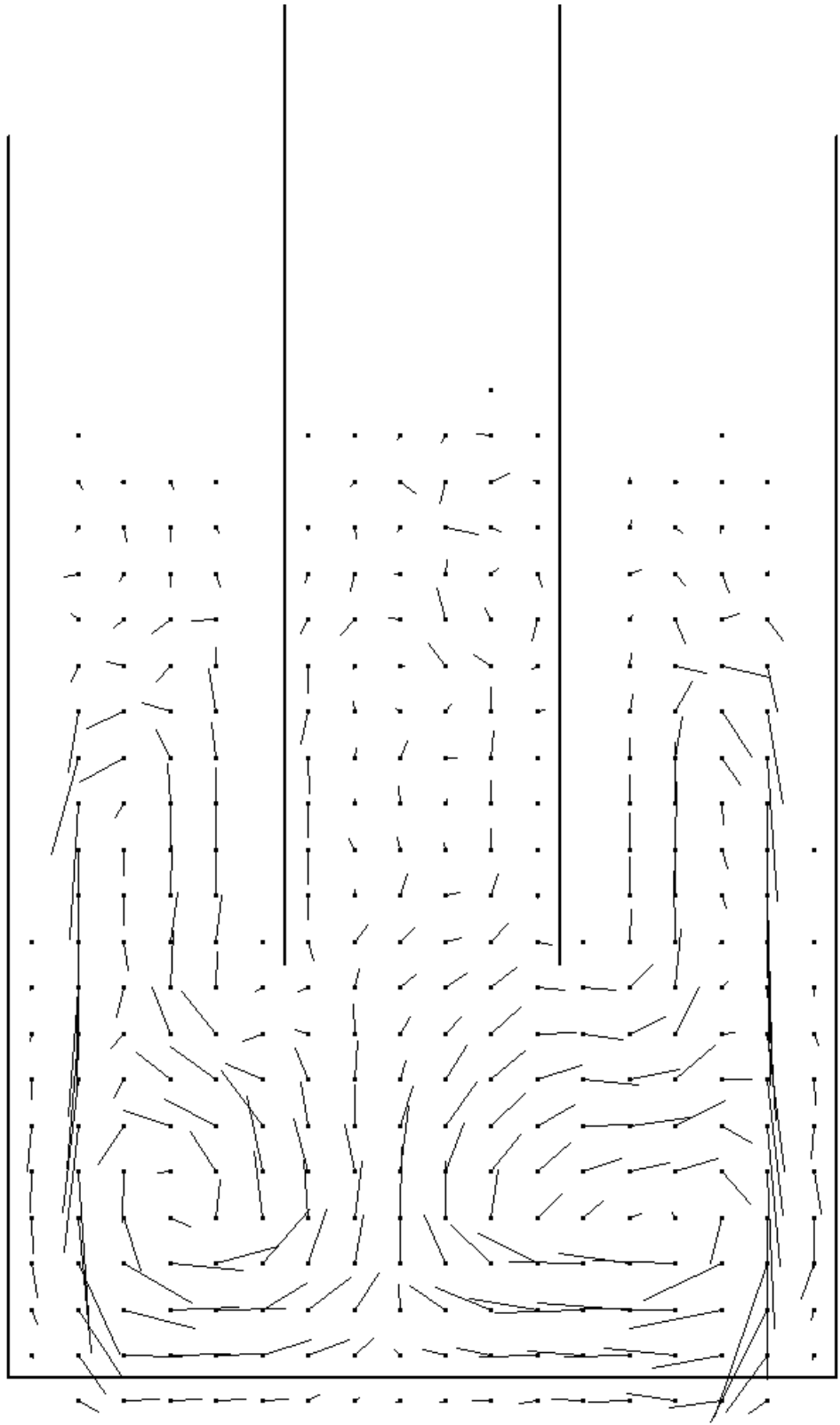
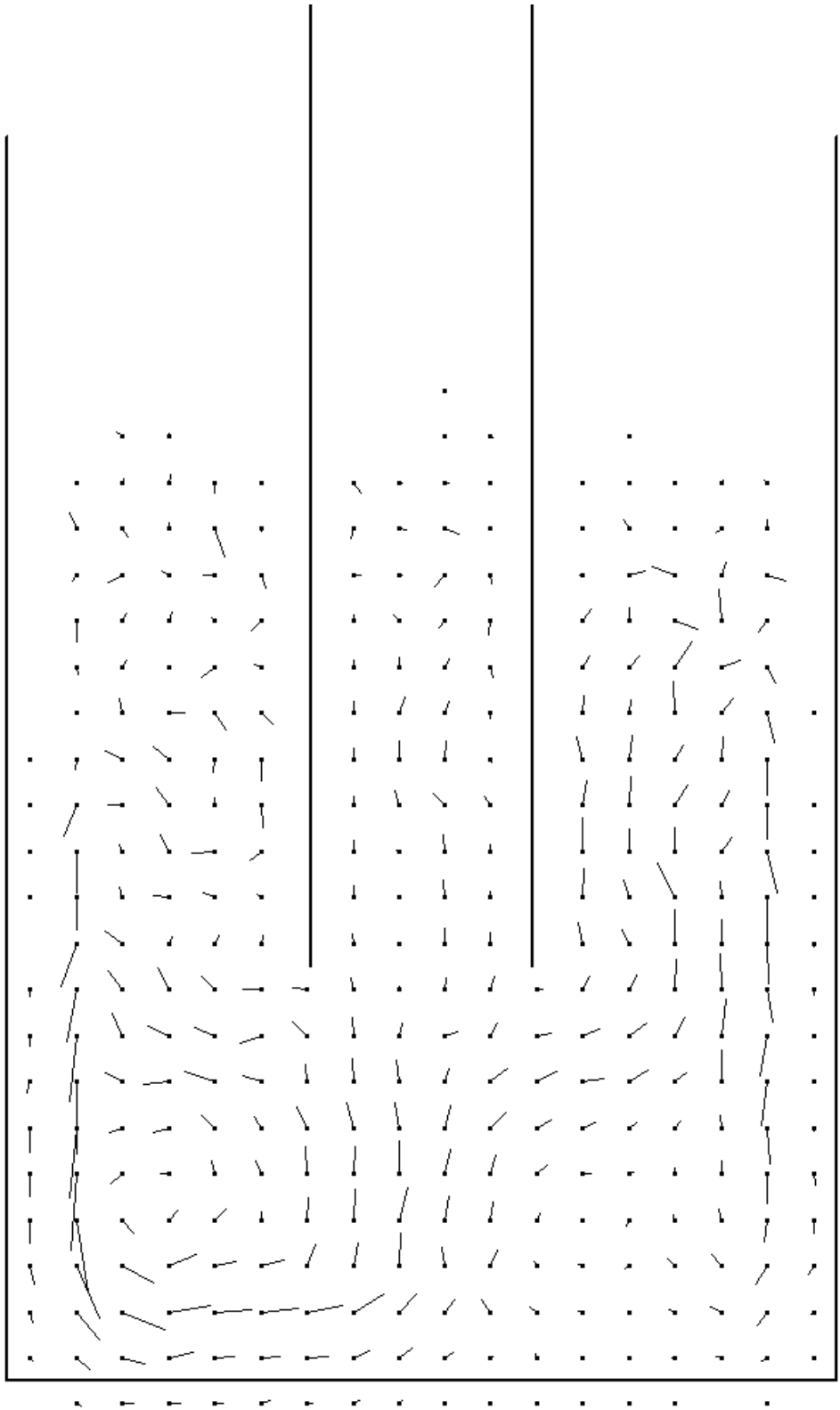
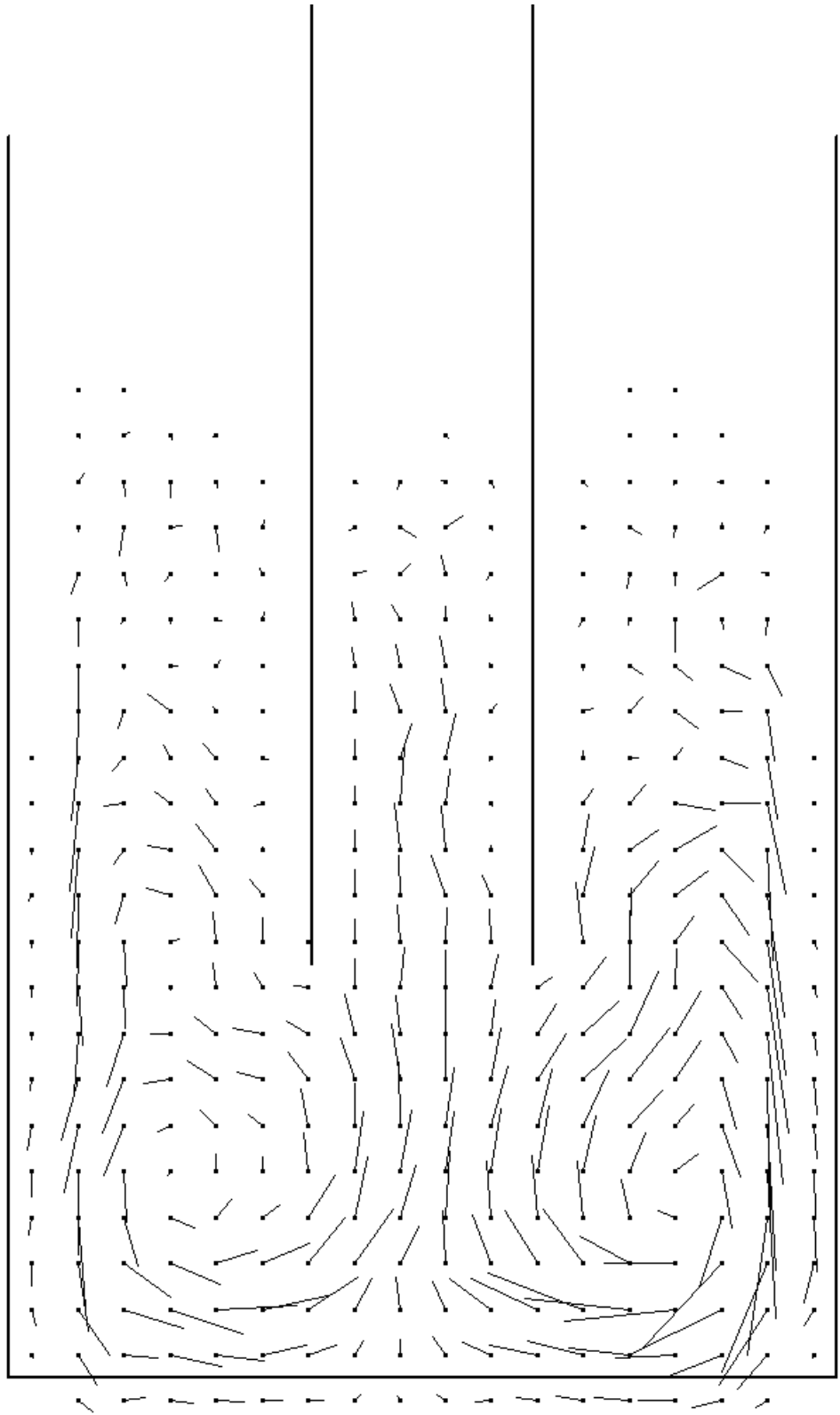
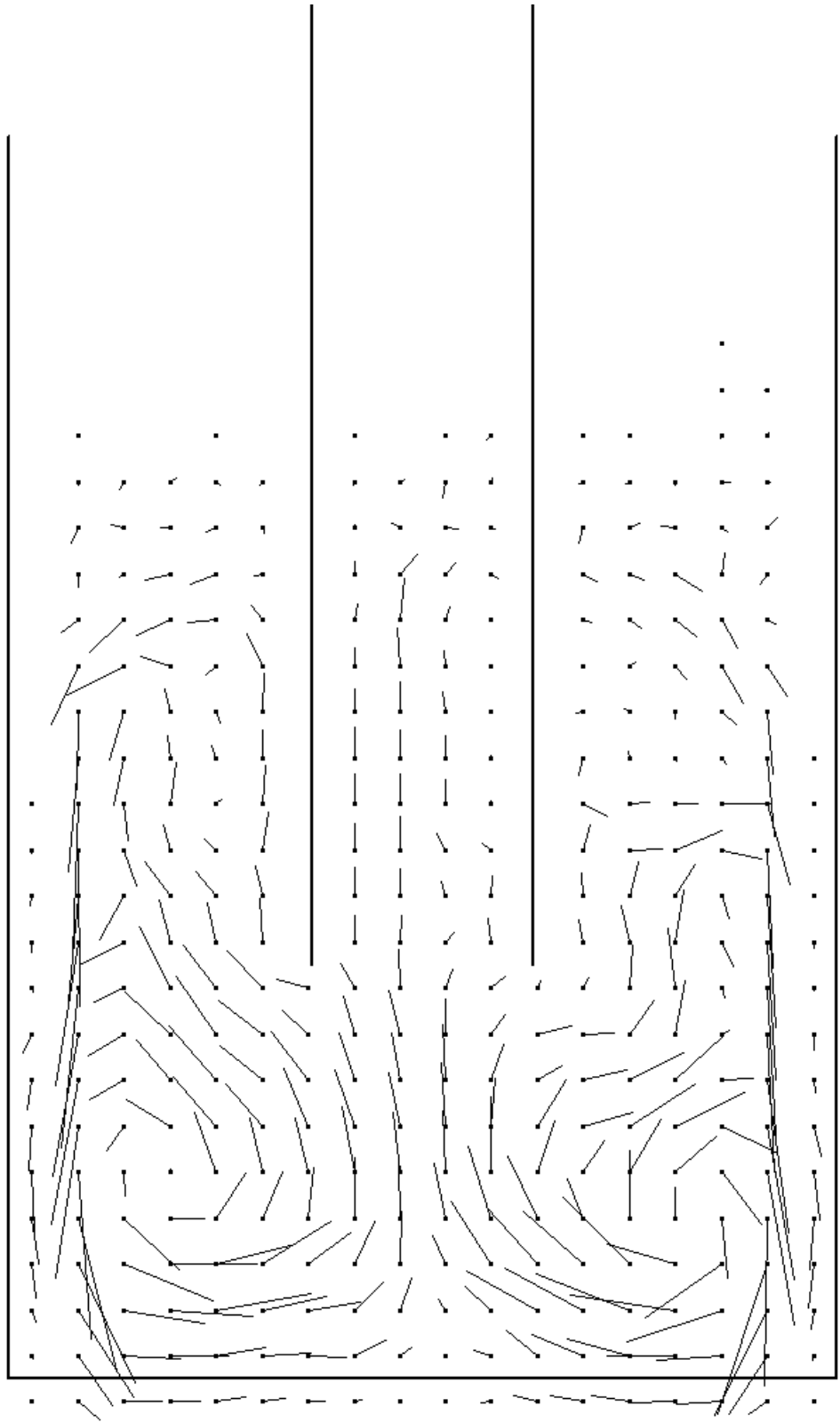
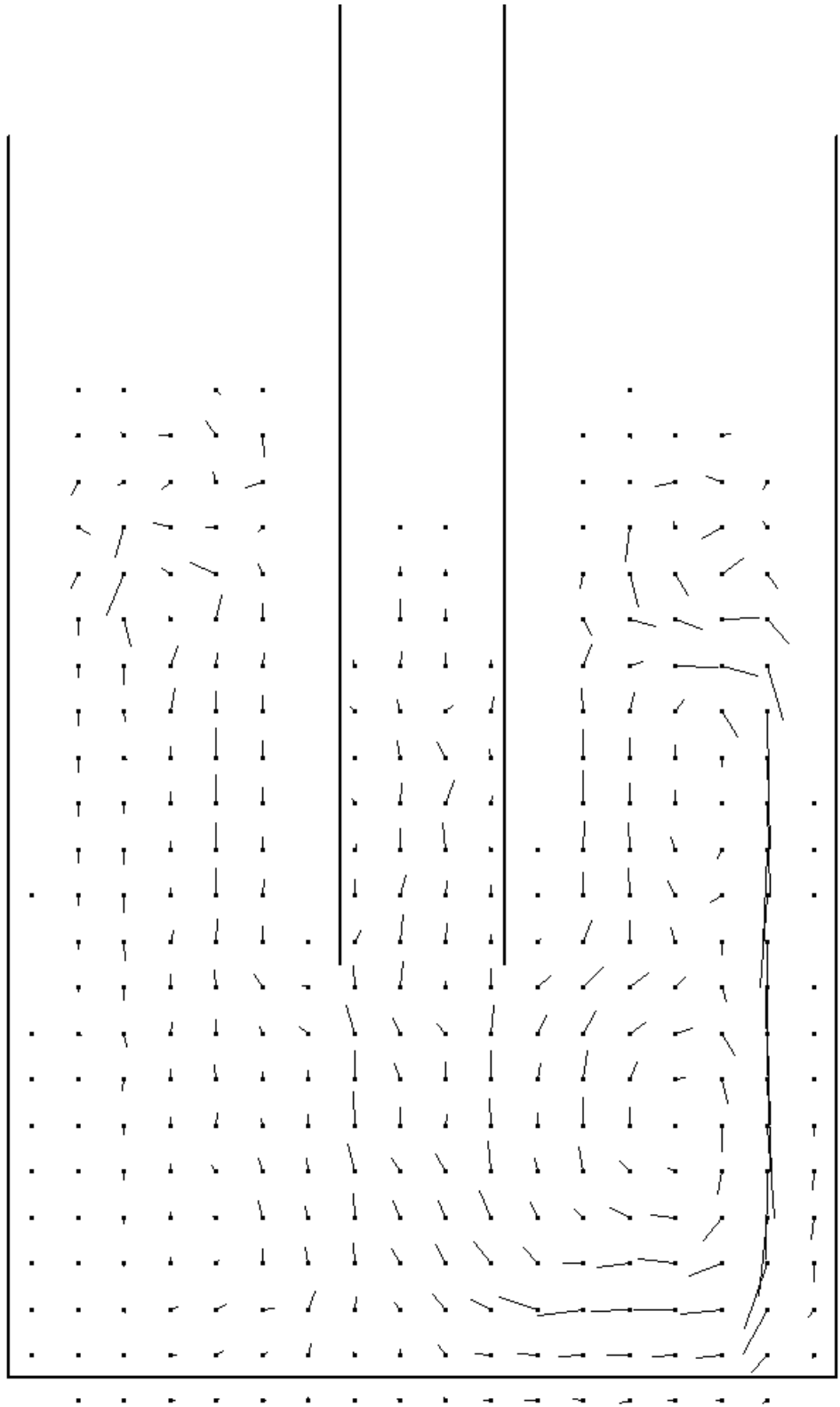
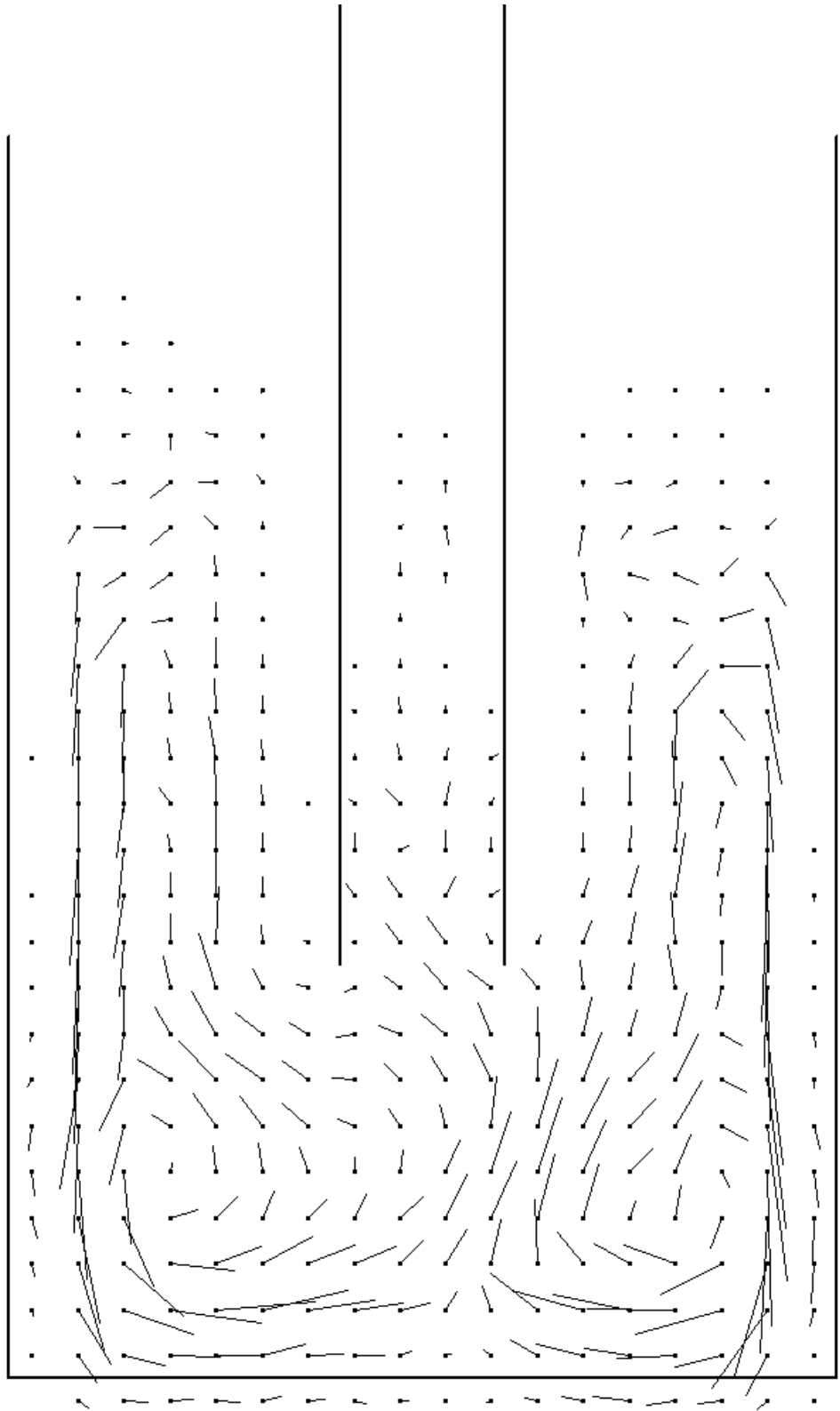
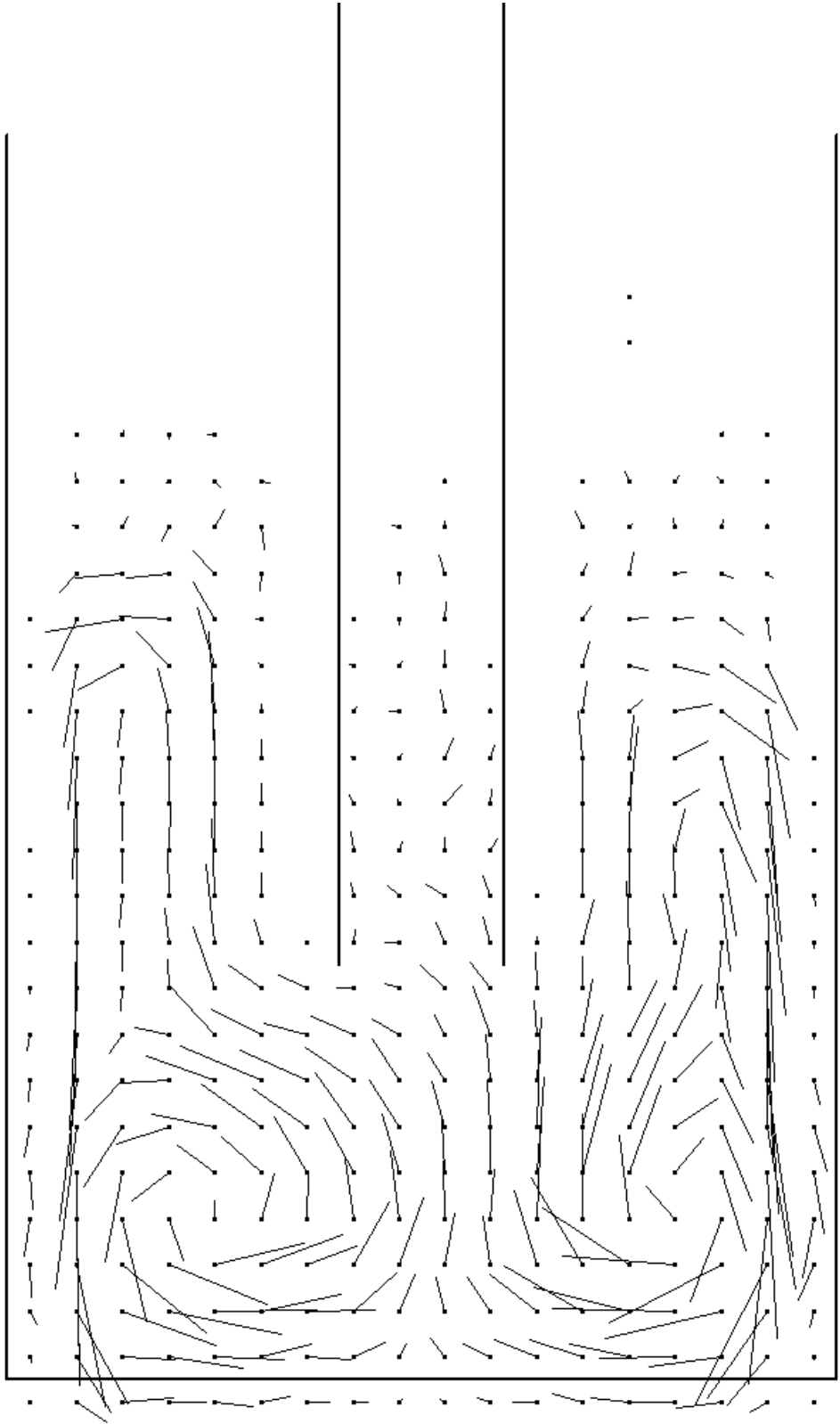
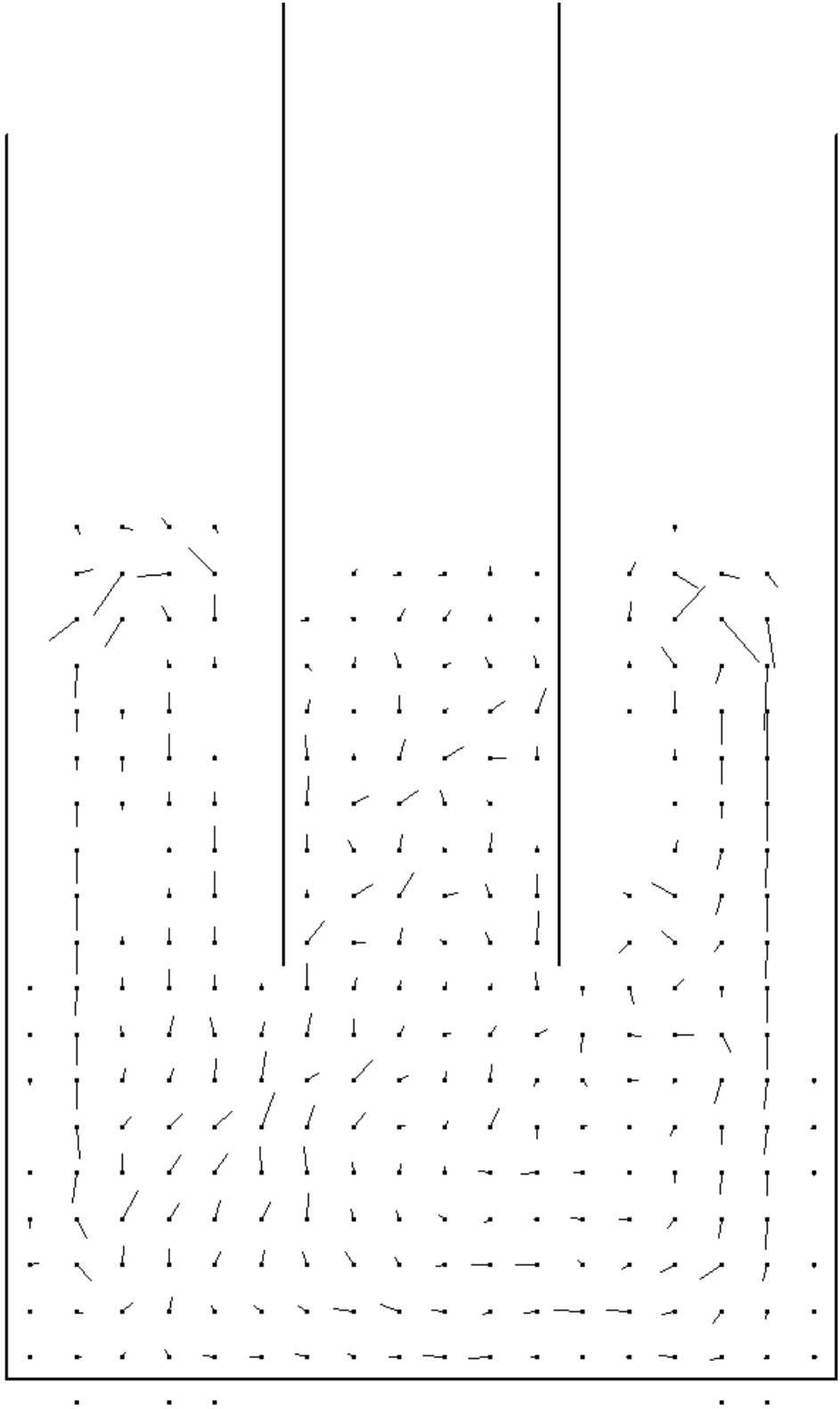
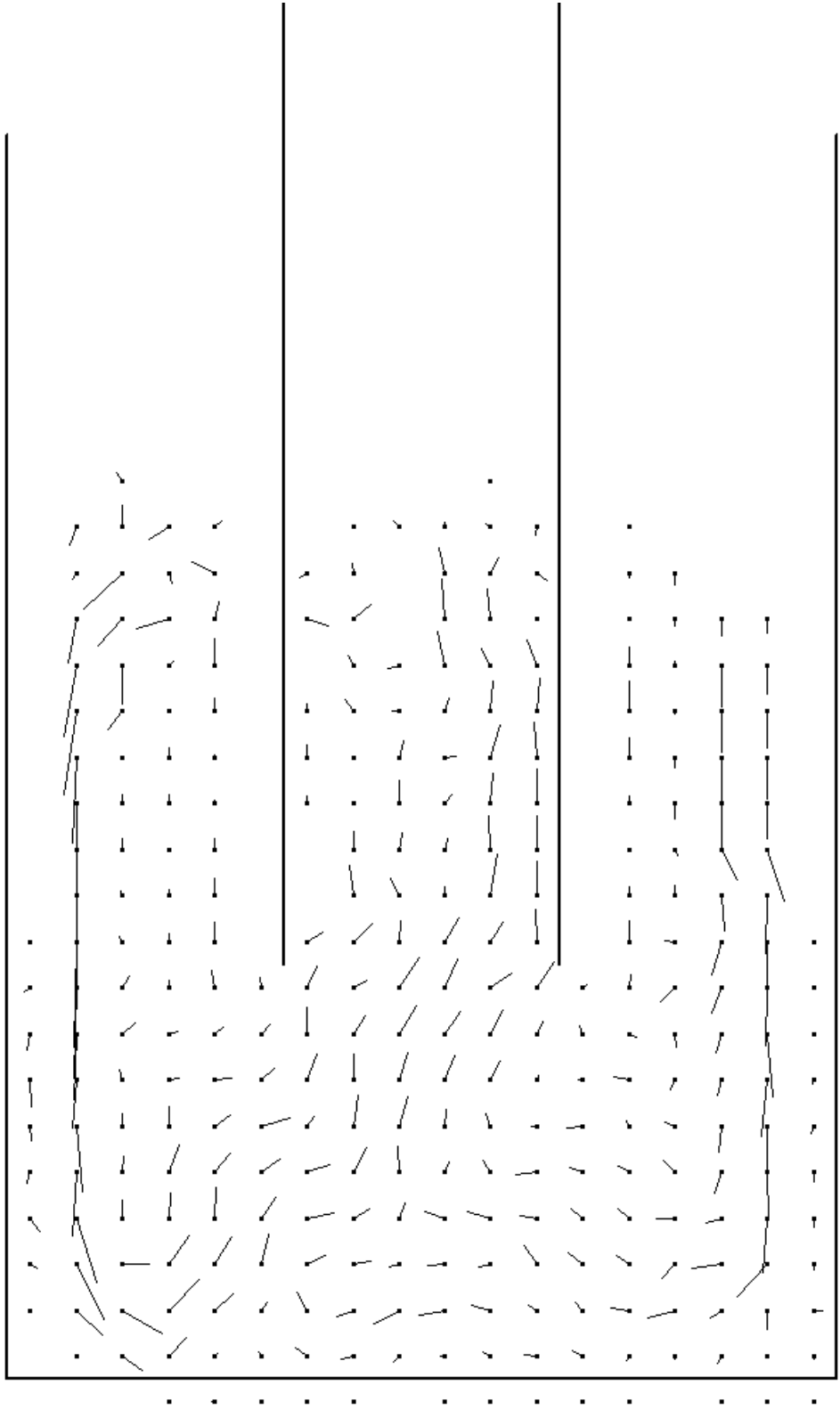
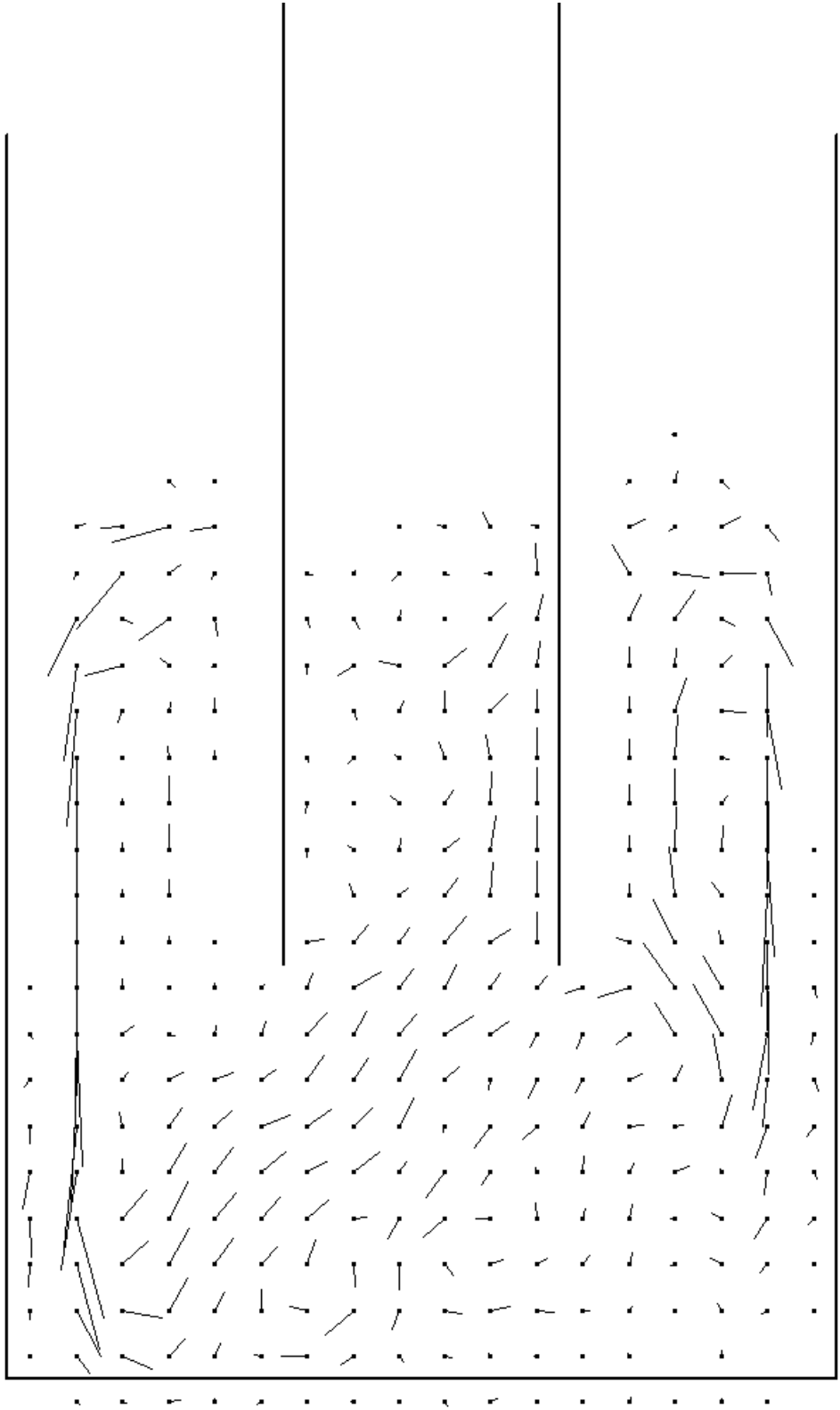
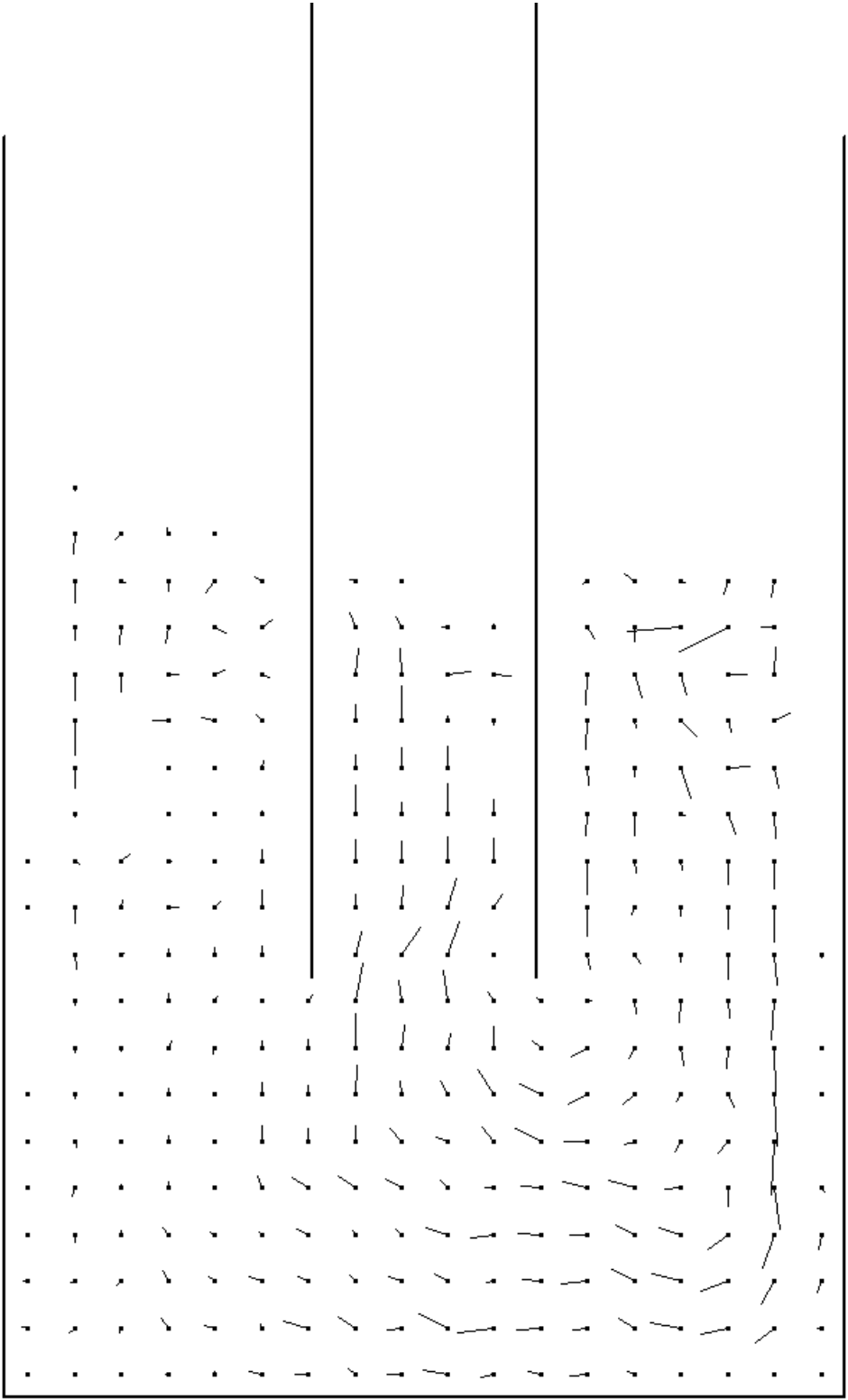
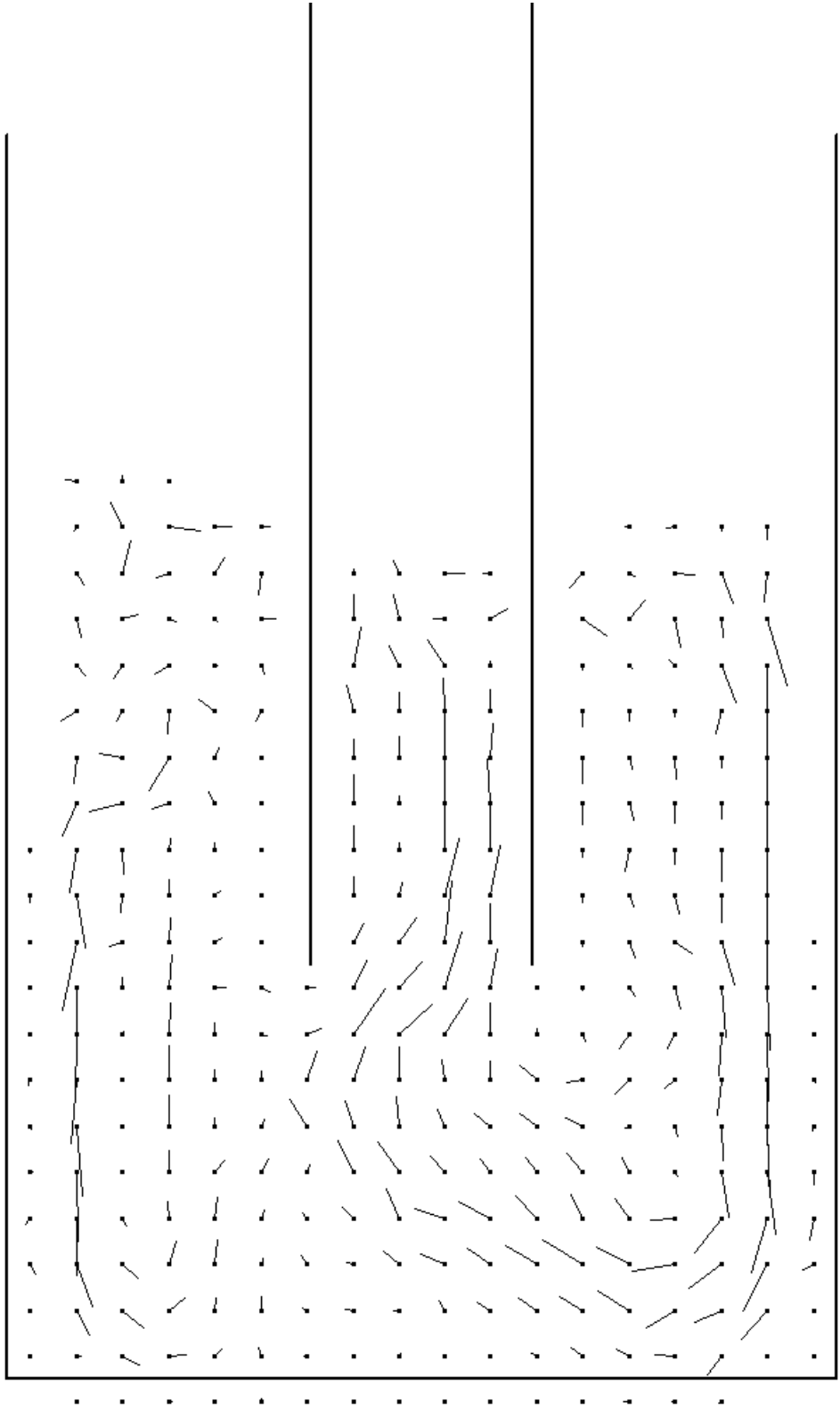
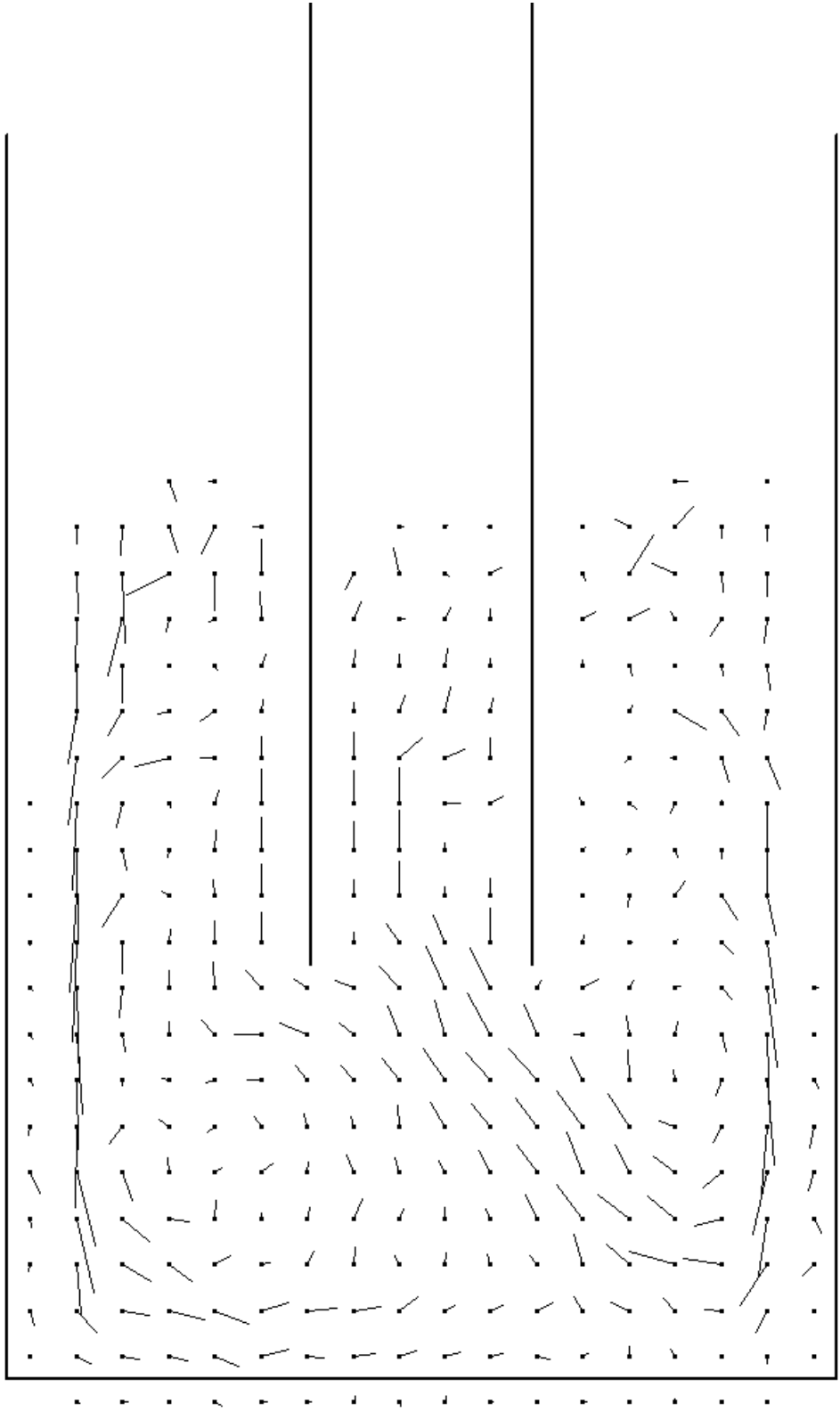
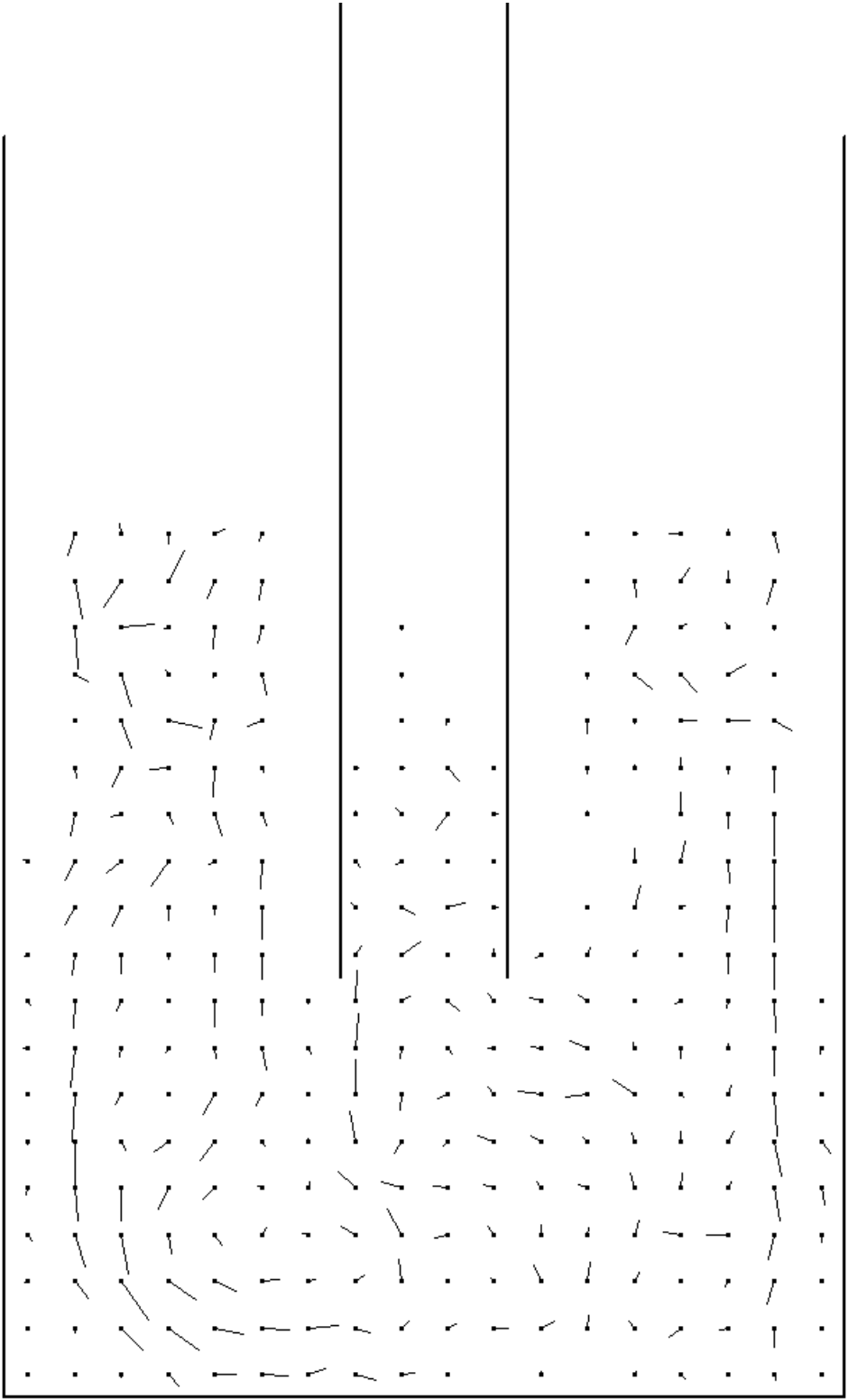
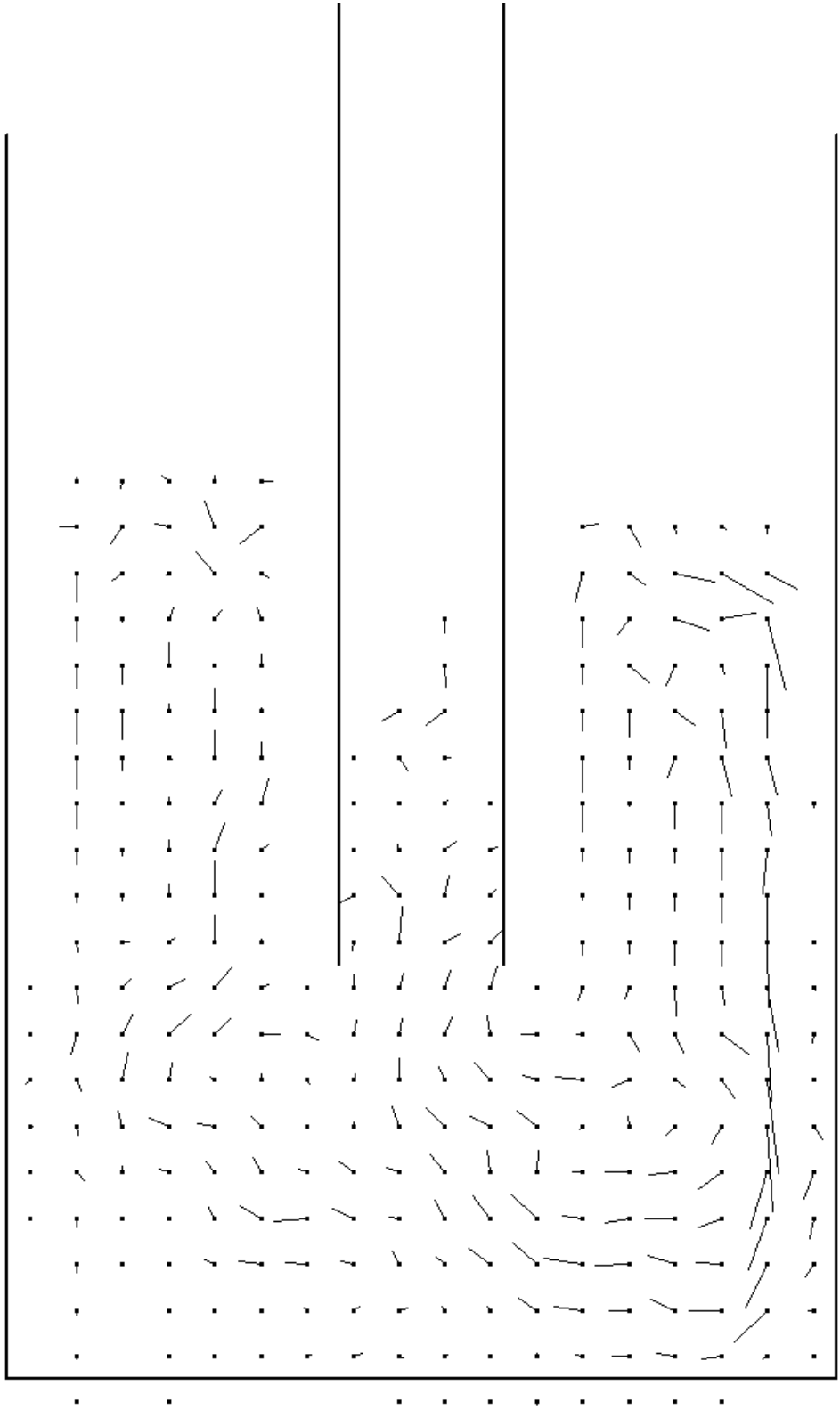
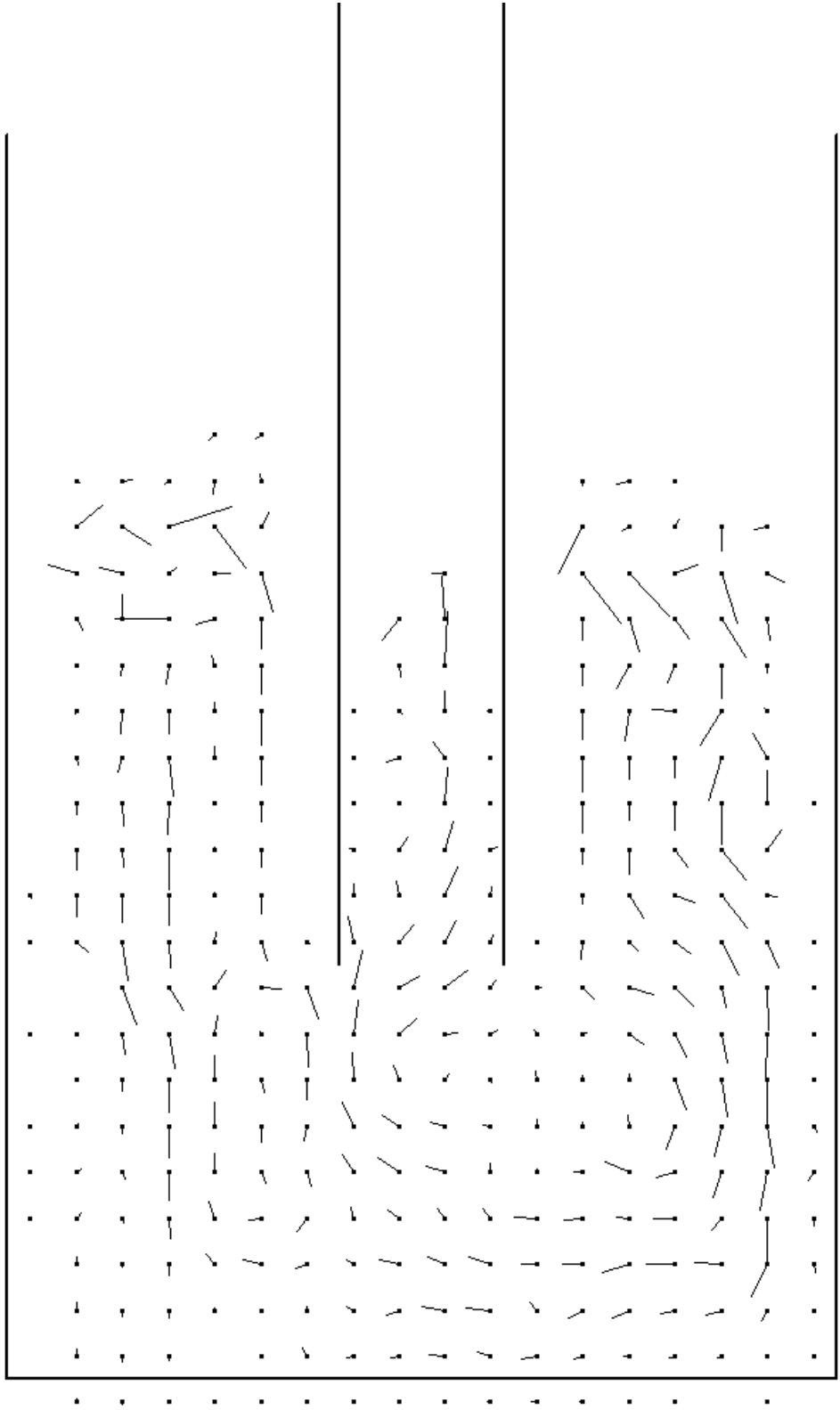
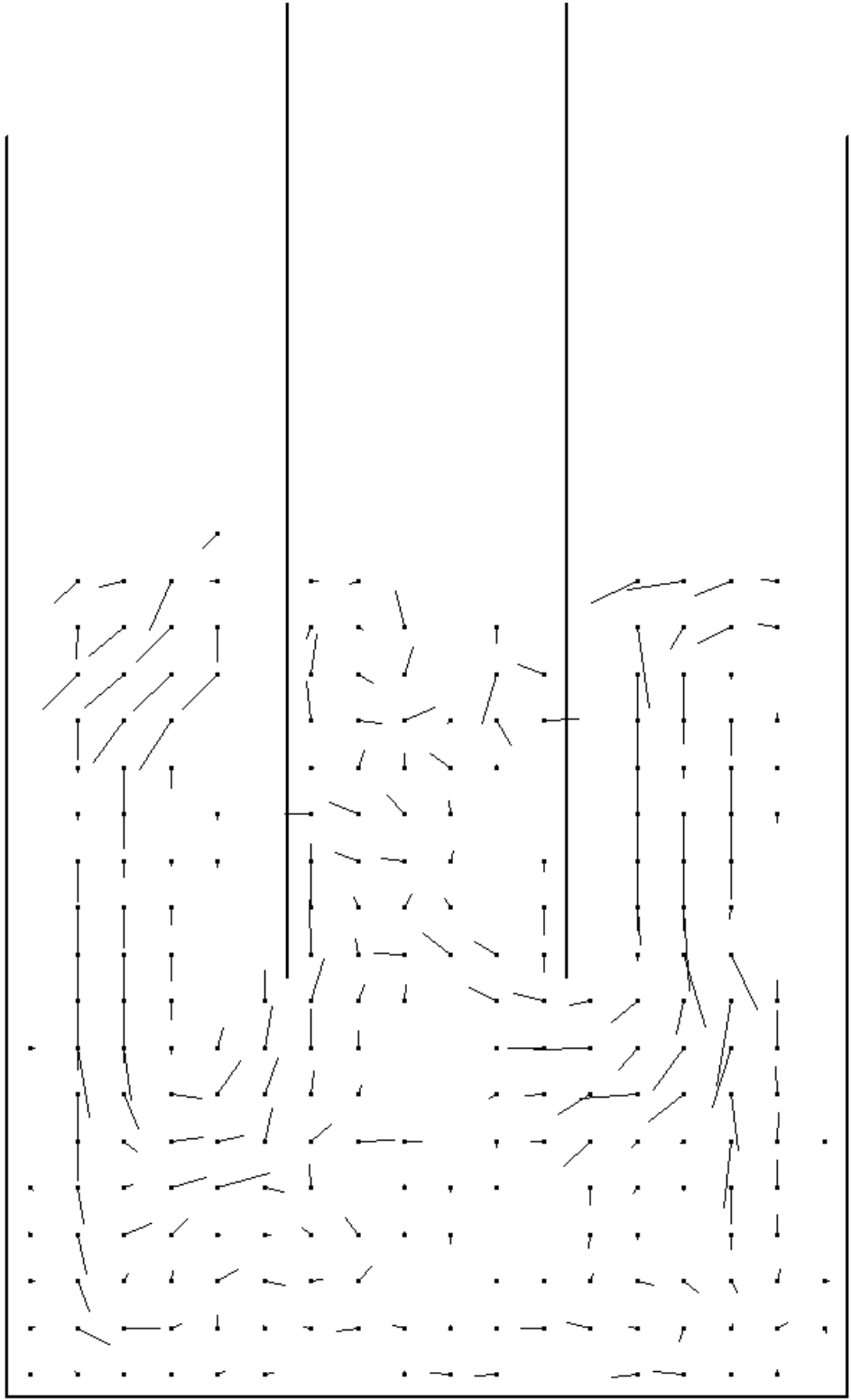
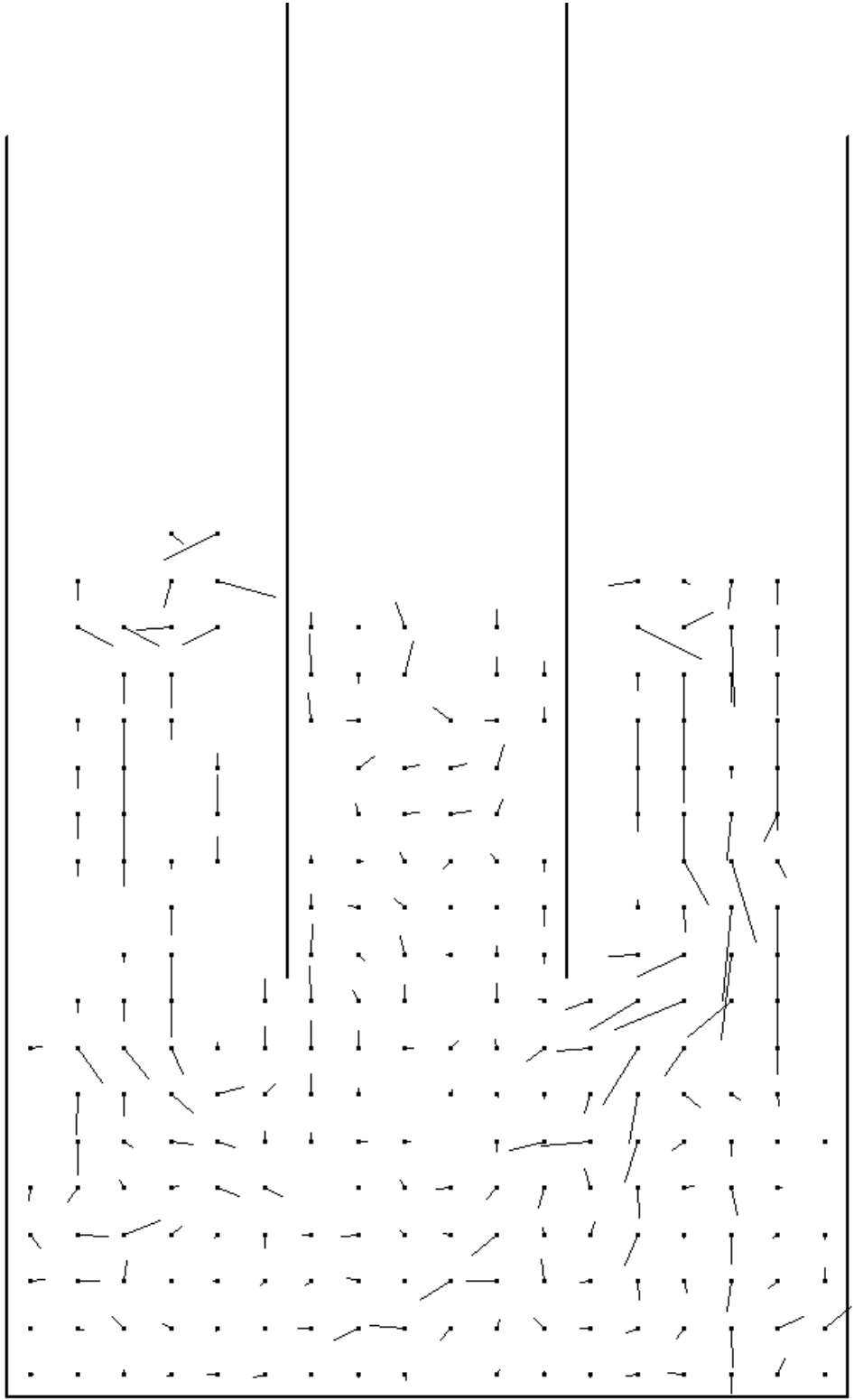
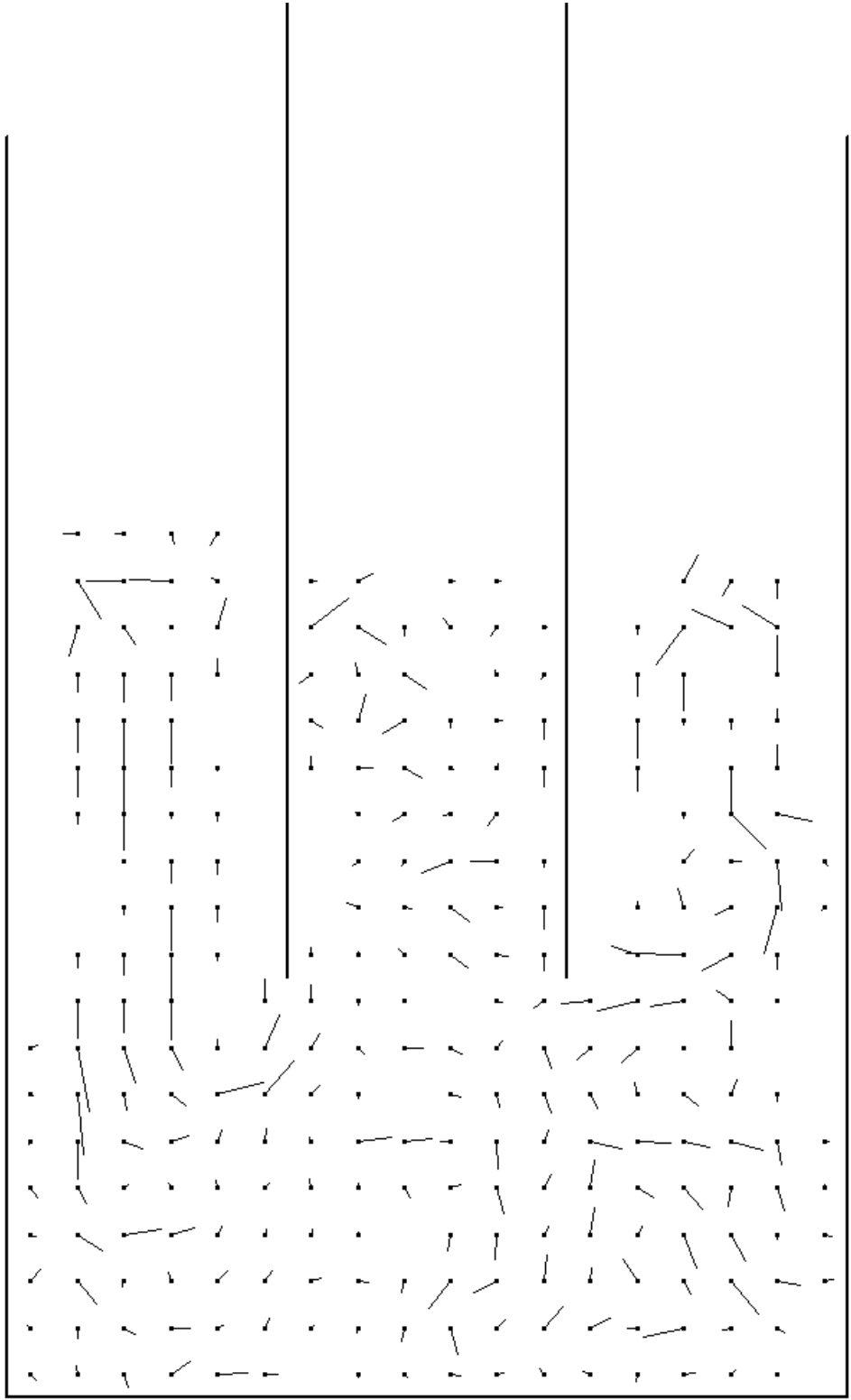
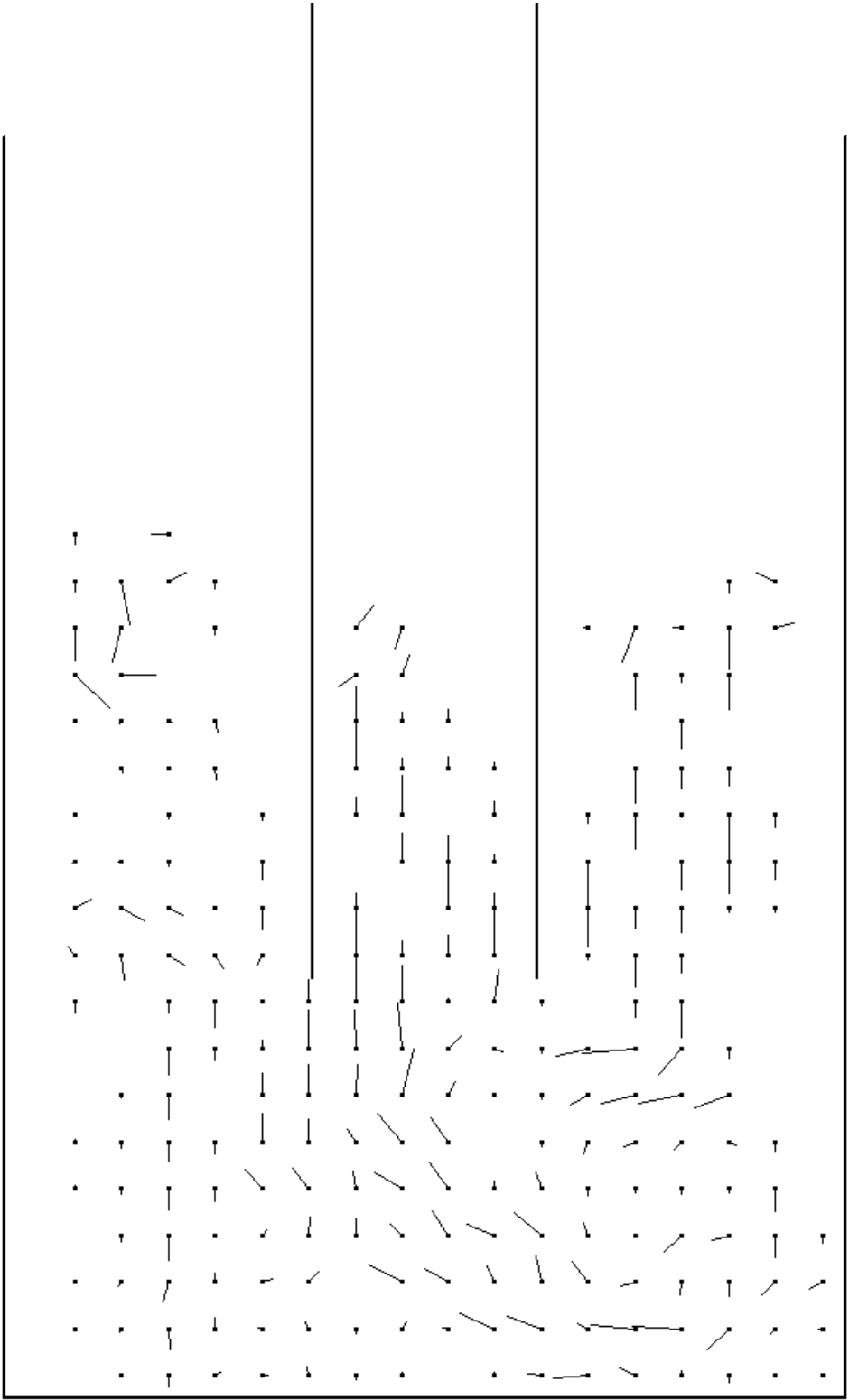
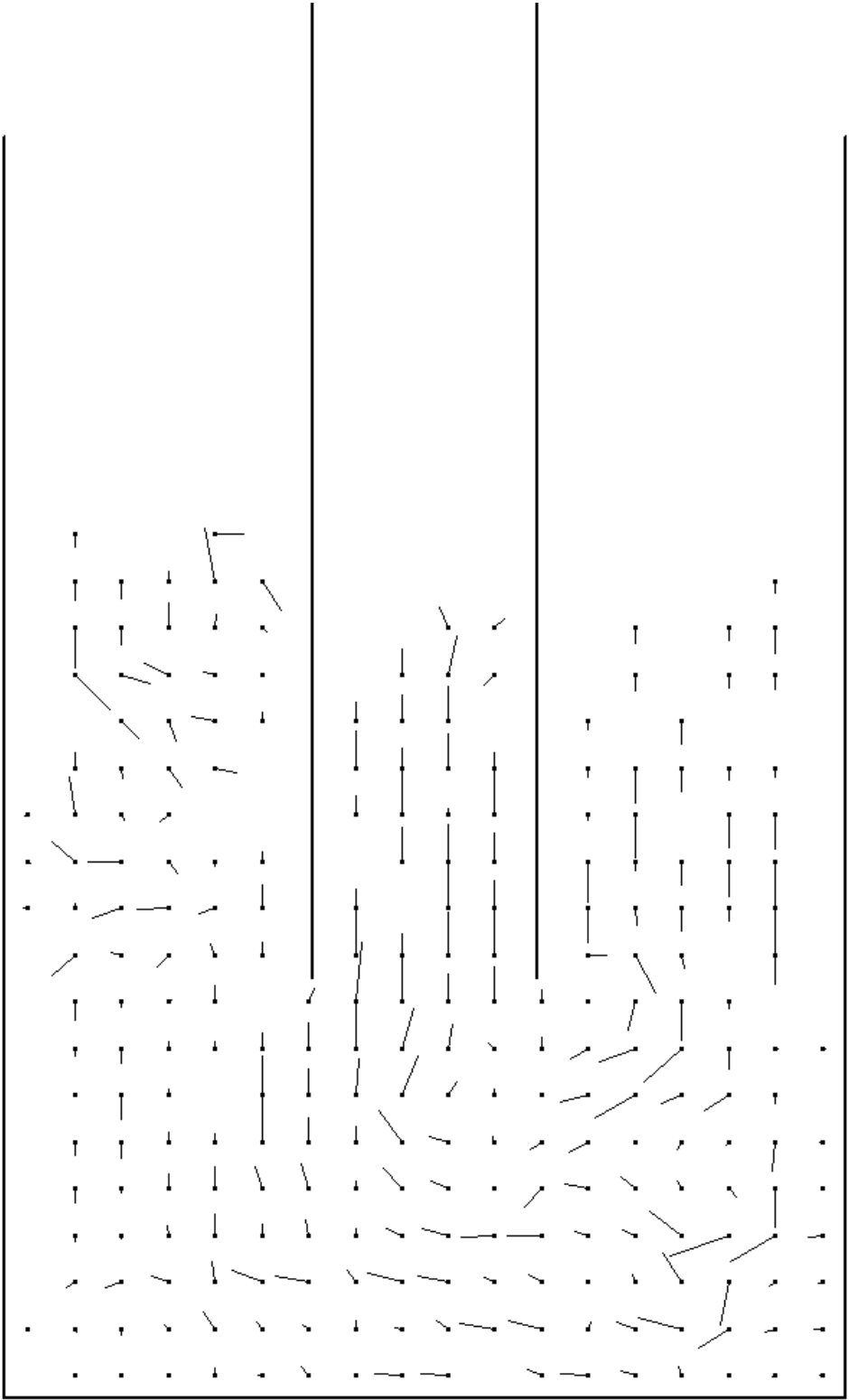
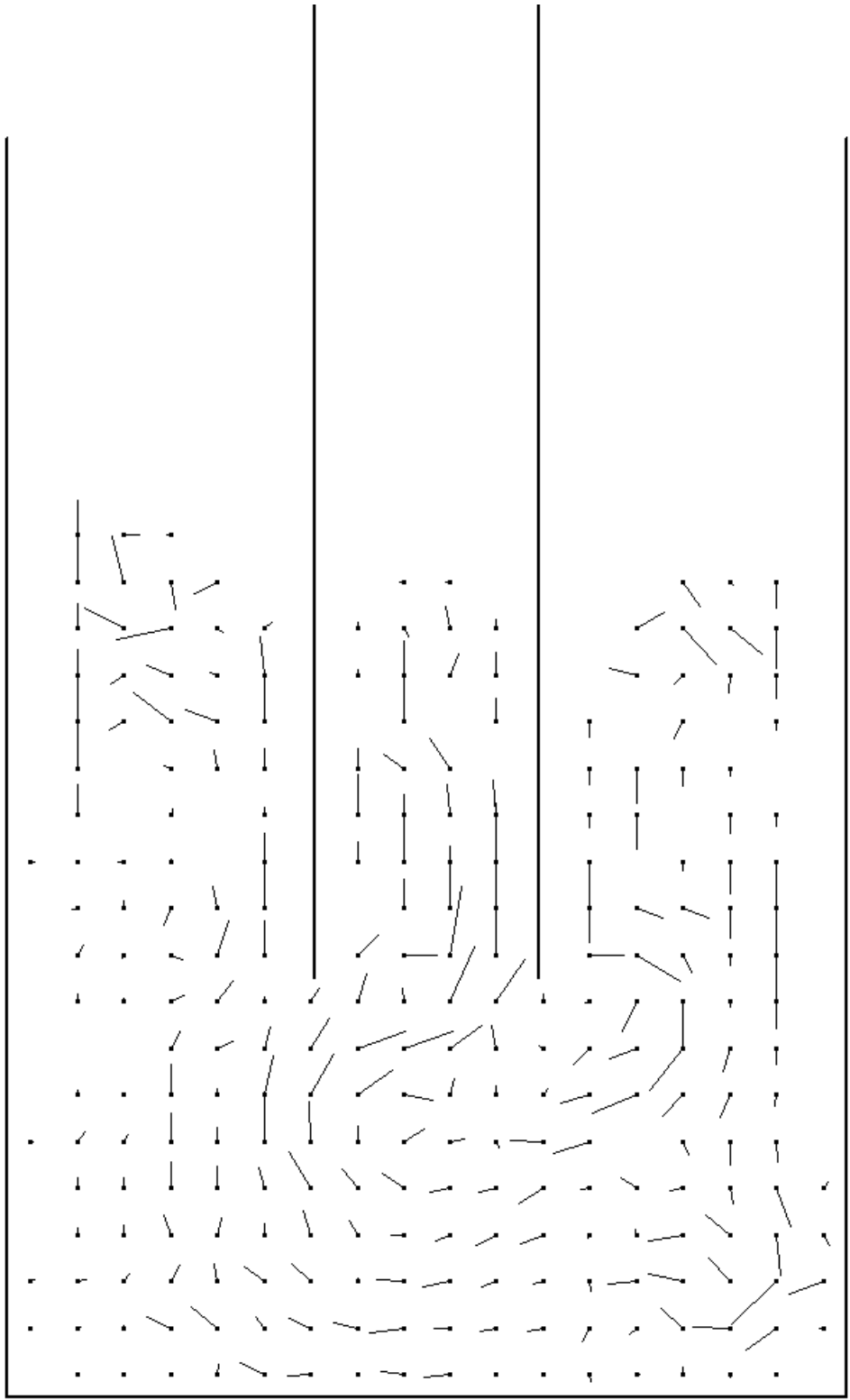
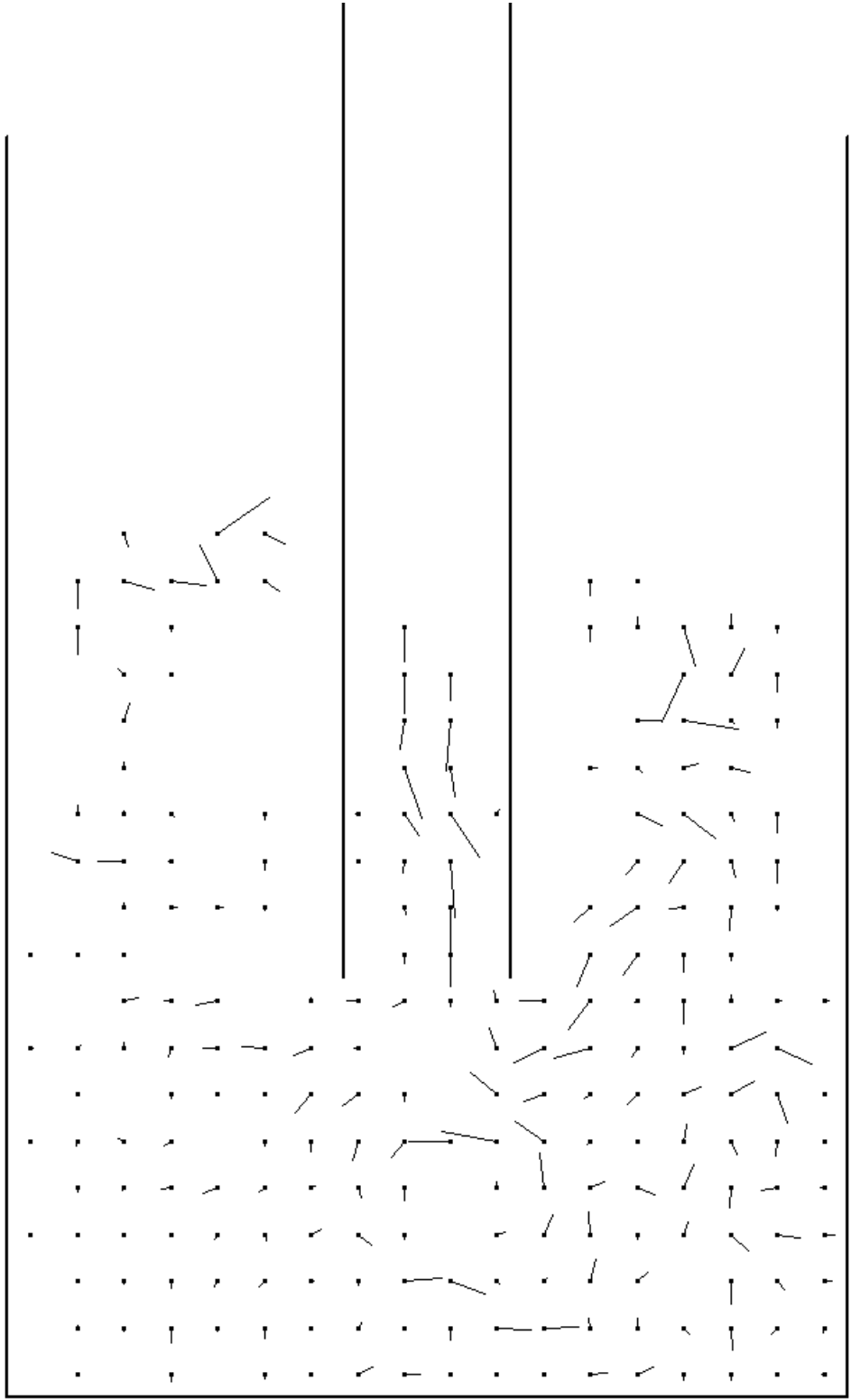
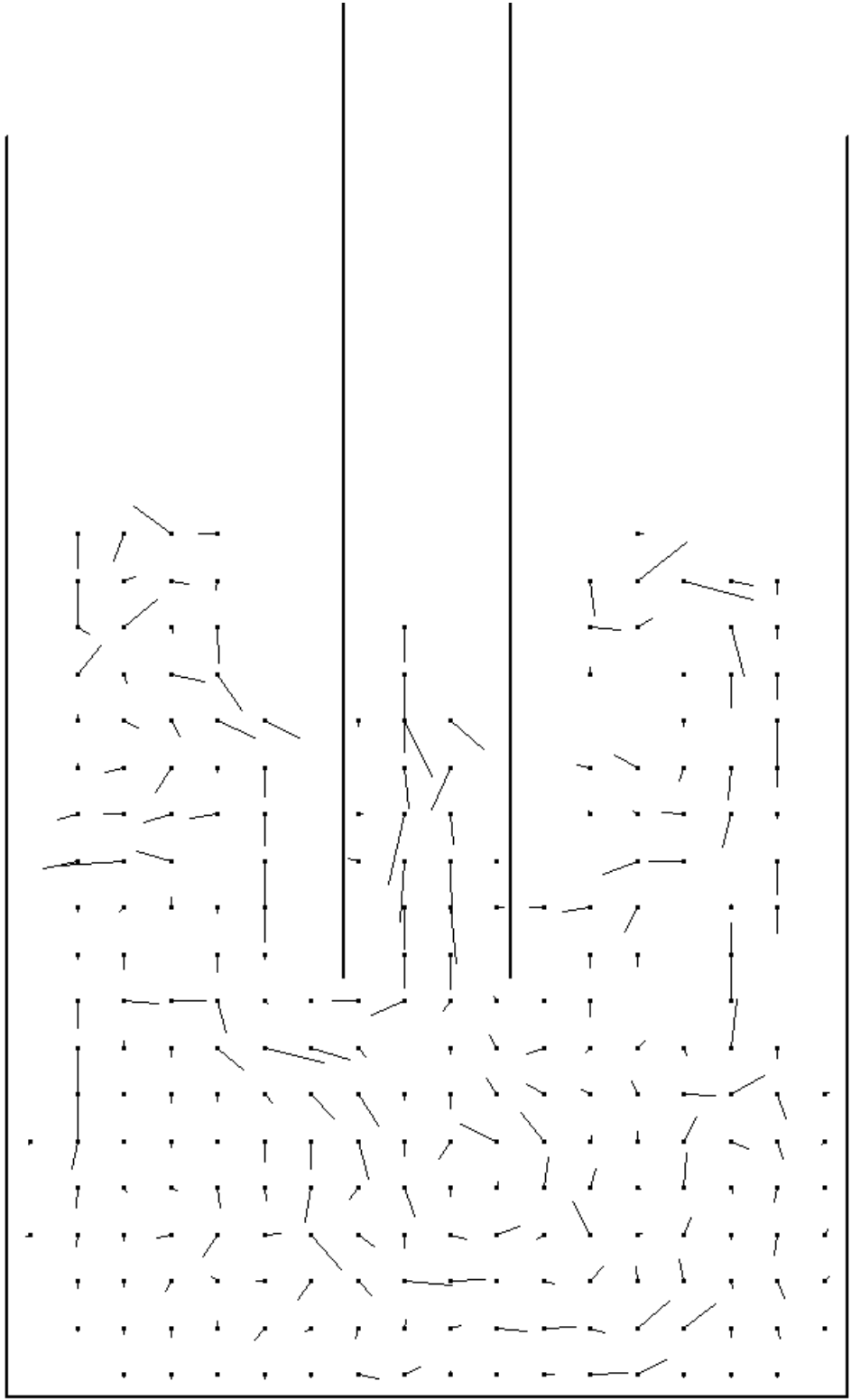
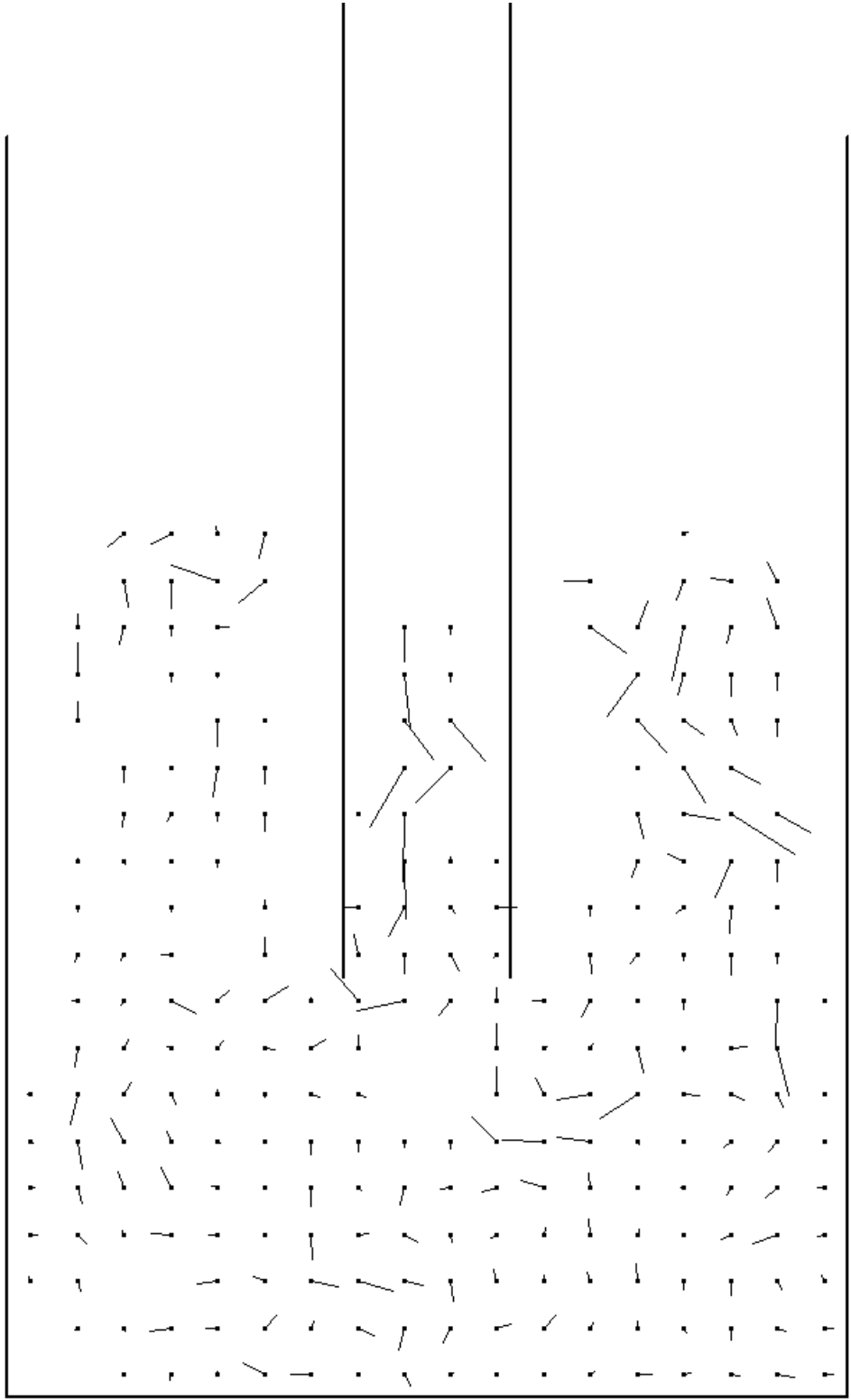
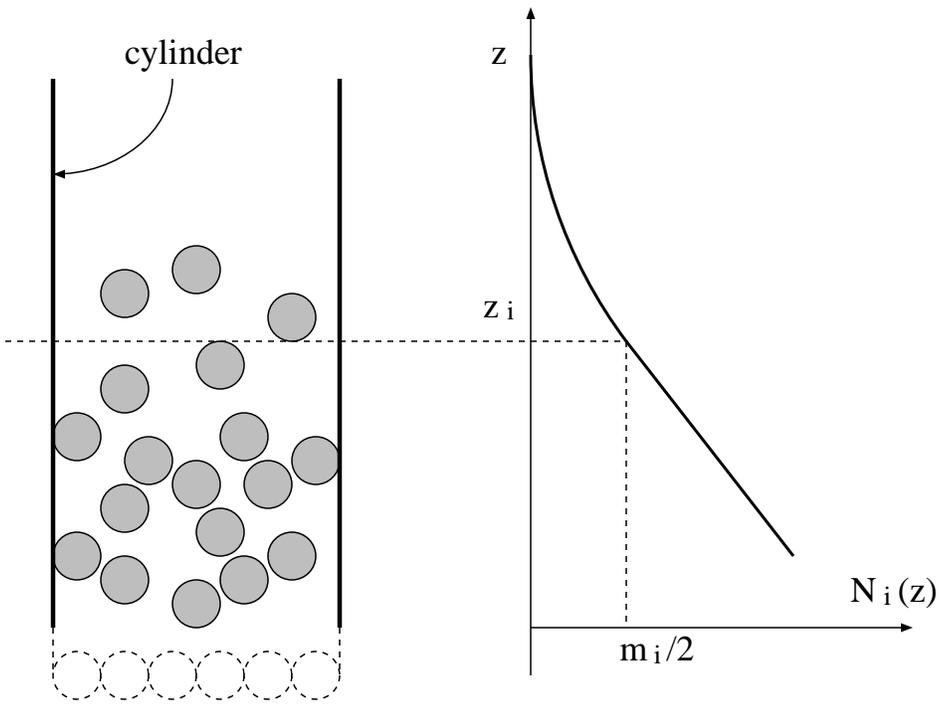
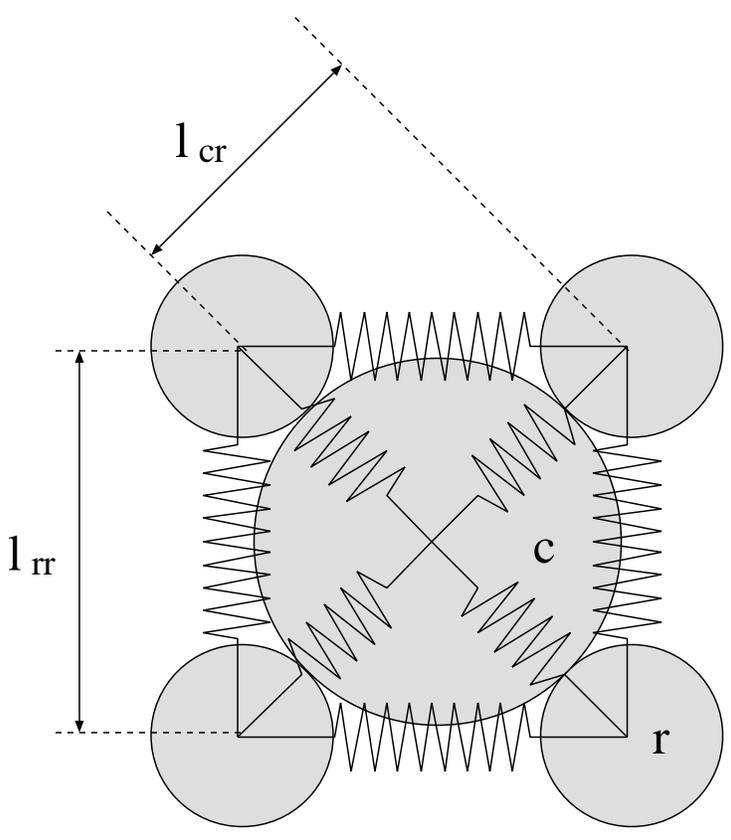
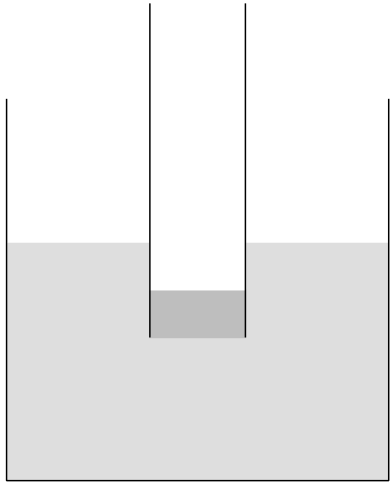
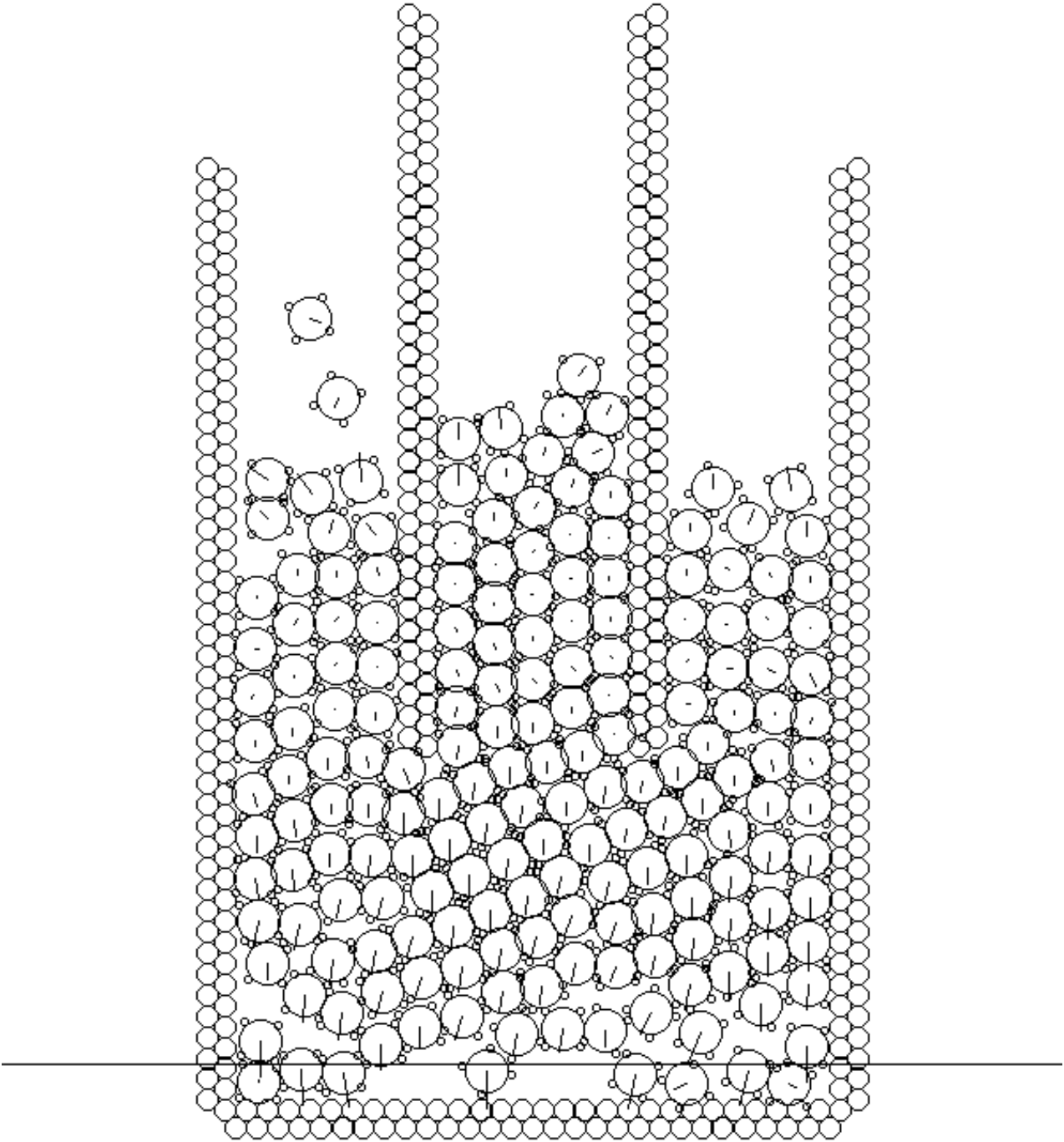
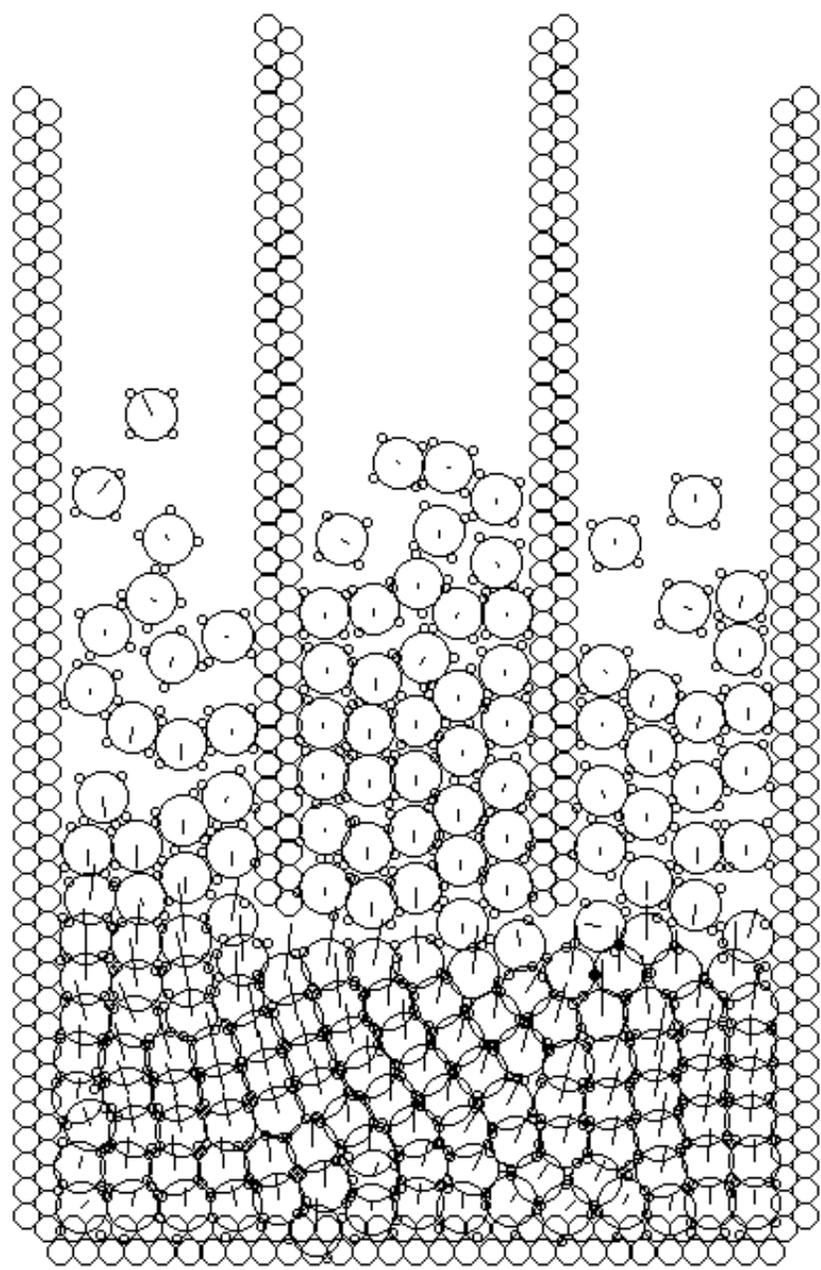
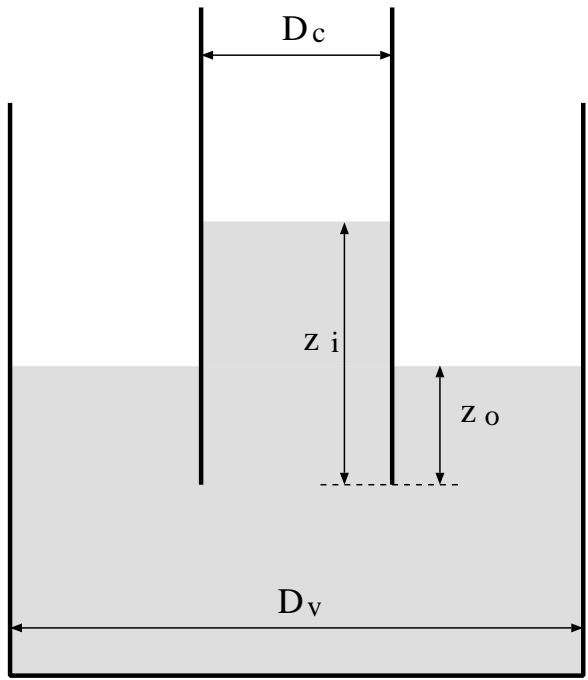
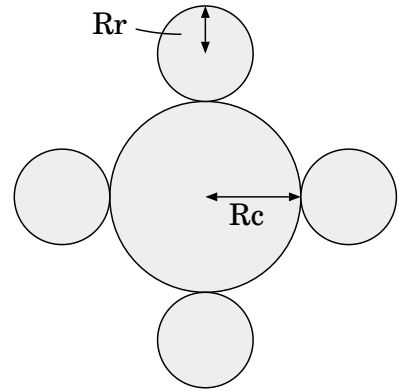

In this paper, we numerically investigated height difference $\Delta z$
between inside and outside injected tube.
$\Delta z$ increases as diameter $D_c$ of injected tube increases
except for a few cases (See Fig. \ref{fig:del z}).
When accelerated amplitude of vibration
$\Gamma$ and $D_c$ are fixed, $\Delta z$ decreases as
angular frequency of vibration $\omega$ increases.
Since convection becomes weak as $\omega$ increases except for a few cases
(See Fig. \ref{fig:del z}),
$\Delta z$ can be regarded as a function of convection indirectly.
These results are schematically shown in Fig.\ref{fig:summary}.
Upper raw corresponds to larger $\omega$ and
lower raw to smaller.
Left column shows the results for smaller $D_c$ and
right column shows those for larger.
Thus, in upper left case, $\Delta z$ takes minimum,
and in lower right case, $\Delta z$ takes maximum.
Also convection which occurs when $\omega$ is small enough
is drawn in lower raw.

Possible qualitative explanation of these results are as follows.
Granular material inside tube is affected by both upward and downward
forces. Upward force is due to convection. Convection is upward
at the center of vessel, thus it push up granular material inside tube.
Difference between upper raw and lower raw is
the difference of strength of convection.
Lower raw has stronger upward force due to convection, thus
surface level inside tube is higher in lower raw than in upper raw.

Downward force is possibly due to difference of density
between inside and outside tube.
Generally, fluidized granular matter has smaller density than
that of fixed bed.
However, granular matter inside tube is hard to flow due to
friction with tube, thus
density is relatively high.
This effect pushes down granular matter into tube.
This conjecture explains why thiner tube has lower level of
granular matter inside tube.
Thiner tube causes higher friction which prevents granular matter
from being fluidized, thus heavier granular matter is hard to
rise by convection.
It is schematically illustrated in Fig.\ref{fig:summary},
where left column has lower level of surface inside tube than
right column.
This conjecture should be confirmed in future.

\section{Acknowledgement}

The author thanks Dr. Hiraku Nishimori at Ibaraki University
for providing his experimental results before publication.
Prof. J. Rajchenbach is also acknowledged for helpful discussion.
The author also thanks Dr. Y-h. Taguchi for his translating Japanese version
of this paper into English and for helpful discussions.

\clearpage
\begin{figure}
\caption{Schematics of experiment\protect\cite{akiyama}
\label{fig:schem}}
\end{figure}
\begin{figure}
\caption{Each non-spherical particle composed of
five sub-particles}
\label{fig:non-spherical}
\end{figure}
\begin{figure}
\caption{Schematics of numerical simulation}
\label{fig:schem.num}
\end{figure}
\begin{figure}
\caption{Snapshot of simulation. Segments attached to each particle
reveals instantaneous velocity vectors}
\label{fig:snapshot}
\end{figure}
\begin{figure}
\caption{Definition of height inside tube.
$N_i(z)$ is number of non-spherical particles above $z$.
The hight $z_i$ inside tube is defined such that $N_i(z)=m_i/2$
where $m_i$ is number of non-spherical particles aligned
along section of tube ($m_i=6$ in this figure).}
\label{fig:height}
\end{figure}
\begin{figure}
\caption{$\Delta z$. $\Diamond : \omega=3.0$,
$+: \omega=4.0$, $\Box: \omega=5.0$.
(a) $\Gamma=5.0$ (b) $\Gamma=6.0$ (c) $\Gamma=7.0$}
\label{fig:del z}
\end{figure}
\begin{figure}
\caption{Flow patterns and its strength ($I$).
(a) $\Gamma=5.0, \omega=3.0$.
Wide tube: $I=1302$, middle tube:$I=1103$, narrow tube:$I=1891$.
(b) $\Gamma=6.0, \omega=3.0$.
Wide tube: $I=2821$, middle tube:$I=2426$, narrow tube:$I=2546$.
(c) $\Gamma=7.0, \omega=3.0$.
Wide tube: $I=2800$, middle tube:$I=2426$, narrow tube:$I=3409$.
(d) $\Gamma=5.0, \omega=4.0$.
Wide tube: $I=446$, middle tube:$I=436$, narrow tube:$I=433$.
(e) $\Gamma=6.0, \omega=4.0$.
Wide tube: $I=686$, middle tube:$I=707$, narrow tube:$I=554$.
(f) $\Gamma=7.0, \omega=4.0$.
Wide tube: $I=931$, middle tube:$I=746$, narrow tube:$I=551$.
(g) $\Gamma=5.0, \omega=5.0$.
Wide tube: $I=368$, middle tube:$I=235$, narrow tube:$I=212$.
(h) $\Gamma=6.0, \omega=5.0$.
Wide tube: $I=311$, middle tube:$I=276$, narrow tube:$I=289$.
(i) $\Gamma=7.0, \omega=5.0$.
Wide tube: $I=257$, middle tube:$I=309$, narrow tube:$I=233$.
}
\label{fig:flow}
\end{figure}
\end{document}